\documentclass[conference,10pt]{IEEEtran}
\IEEEoverridecommandlockouts
\usepackage{graphicx,dblfloatfix}
\usepackage{cite}

\usepackage{amsmath,amssymb,amsfonts}
\usepackage{algorithmic}
\usepackage{verbatim}
\usepackage{graphicx}
\usepackage{textcomp}
\usepackage{xcolor}
\usepackage{lipsum}
\usepackage{subcaption}
\usepackage{wrapfig}
\usepackage{ulem}[normalem]
\usepackage{bm}
\usepackage{enumitem,kantlipsum}
\usepackage{booktabs}
\usepackage{adjustbox}
\usepackage{url}

\usepackage{multirow}
\def\BibTeX{{\rm B\kern-.05em{\sc i\kern-.025em b}\kern-.08em
    T\kern-.1667em\lower.7ex\hbox{E}\kern-.125emX}}
\DeclareMathOperator{\ricK}{\mathcal{K}}
\DeclareMathOperator{\fftp}{\mathcal{FFT}_P}
\DeclareMathOperator{\ifftp}{\mathcal{IFFT}_P}

\DeclareMathOperator{\expct}{\mathbb{E}_P}
\newcommand{\argmax}[1]{\underset{#1}{\operatorname{arg}\,\operatorname{max}}\;}
\usepackage{pifont}
\newcommand{\xmark}{\ding{55}}%
\begin{document}
\title{\huge Adaptive High-Speed Radar Signal Processing Architecture for 3D Localization of Multiple Targets on System on Chip}
\author{
\IEEEauthorblockN{}
\IEEEauthorblockA{}
}

\author{Aakanksha Tewari,  Jai Mangal, Sumit~J~Darak~\IEEEmembership{Senior Member~IEEE},\\ Shobha~Sundar~Ram,~\IEEEmembership{Senior Member~IEEE}, and Arnav Shukla 
\thanks{The authors would like to thank Ankit Kumar Pal (B.Tech., 2025, IIIT Delhi) and Shragvi Sidharth Jha (B.Tech., 2023, IIIT Delhi) for their insightful discussions and inputs.}
\thanks{All authors are with the Indraprastha Institute of Information Technology Delhi, New Delhi 110020 India. E-mail: \{aakankshat,jaim,sumit, shobha,arnav22103\}@iiitd.ac.in.}}

\maketitle
\begin{abstract}
Integrated Sensing and Communication (ISAC) is a key enabler of high-speed, ultra-low-latency vehicular communication in 6G. ISAC leverages radar signal processing (RSP) to localize multiple unknown targets amid static clutter by jointly estimating range, azimuth, and Doppler velocity (3D), thereby enabling highly directional beamforming toward intended mobile users. However, the speed and accuracy of RSP significantly impact communication throughput.
This work proposes a novel 3D reconfigurable RSP accelerator, implemented on a Zynq Multi-processor System-on-Chip (MPSoC) using a hardware–software co-design approach and fixed-point optimization. We propose two RSP frameworks: (1) high-accuracy and high-complexity, and (2) low-complexity and low-accuracy, along with their respective architectures. Then, we develop an adaptive architecture that dynamically switches between these two frameworks based on the signal-to-clutter-plus-noise ratio. This adaptive reconfiguration achieves up to 5.6$\times$ faster RSP compared to state-of-the-art designs. At the system level, the proposed RSP-based ISAC delivers a 24\% improvement in communication throughput without increasing hardware complexity.

\end{abstract}


\section{Introduction}
Integrated Sensing and Communication (ISAC) at millimeter-wave (mmWave) frequencies is an emerging and significant focus area in 6G standardization \cite{kvm2019mmwjrc, RSP_hw4,ISAC_Ieee_proc}. ISAC requires accurate and fast radar signal processing (RSP) to detect mobile users (MUs) within a wireless environment and utilize the sensed information for applications such as high-speed vehicular communication, digital twin networks, and extended reality (XR). This has led to various works focusing on efficient algorithms and architectures for RSP to estimate the range, azimuth, and Doppler velocity of multiple targets \cite{RSP_hw1,RSP_hw2,CAS1_beamtrain, CAS1_beamtrain2,RSP_hw0,RSP_hw3}.  For instance, the IEEE 802.11ad standard uses beam refinement fields (BRF) within the communication packets and conventional energy detection to enable beam alignment between the base station (BS) and MU \cite{CAS1_80211ad,noauthor_ieee_2016-1}. Recent works have studied the modification of the standard IEEE 802.11ad by replacing BRF with RSP \cite{sneh2022ieee, MAB_JSTSP,9737357}. In such systems, the performance gain—specifically the speed of beam alignment and communication throughput—is highly dependent on the RSP's speed and accuracy relative to BRF. Furthermore, using range-Doppler processing in standalone radars is unsuitable, since ISAC requires radar and communication to share the same hardware and spectrum in a time-multiplexed manner. Consequently, the efficient design of RSP algorithms, their optimized implementation on hardware, and seamless integration with the communication transceiver represent critical research challenges \cite{RSP_hw1,RSP_hw2,RSP_hw0,RSP_hw3}.

The traditional approach in radar systems is to implement range-Doppler processing to separate targets/MUs from clutter, followed by the angle-of-arrival estimation of the target, \cite{RSP_hw0,RSP_hw3,engels2021autoradar,hakob2019ieeespsmagazine}. However, accurate Doppler estimation requires a long coherent processing interval (CPI). When the hardware resources are time-shared between the radar and communication functionalities in ISAC, the long CPI followed by RSP time reduces the available communication time with MU and the overall throughput. A very short CPI, on the other hand, reduces the overall probability of detection of MUs. Therefore, the CPI and RSP times determine the overall detection of MUs and their corresponding communication service times \cite{ram2022optimization}. Hence, several recent works have focused on architectures that reduce the overall computational time and complexity of the RSP \cite{RSP_hw0,RSP_hw3,MUSIC_HW1, li2022musictcas1}. For example, \cite{RSP_hw0,RSP_hw3} considers fast Fourier transform (FFT) based RSP, while \cite{MUSIC_HW1, li2022musictcas1} presents Multiple Signal Classification (MUSIC)-based angle-of-arrival (AoA) estimation. However, they do not consider multiple-target scenarios or account for ISAC constraints while jointly estimating range, azimuth, and Doppler velocity.  
Our earlier work in \cite{tewari2024reconfigrsp} employed joint azimuth-range processing (JARP) via digital beamforming and matched filtering on the first pulse to rapidly localize MU, followed by the Doppler velocity estimation through MUSIC using multiple subsequent pulses. The limitation of \cite{tewari2024reconfigrsp} is that the JARP framework carries high computational complexity and latency. In this work, we present a \textit{modified} JARP (MJARP), with multiple hardware architecture innovations to accelerate JARP's performance for 3D localizations of multiple MUs. We also develop a low-computationally complex sequential azimuth-range processing (SARP) architecture, suitable for high signal-to-clutter-and noise ratio (SCNR) conditions. Furthermore, we estimate the Doppler velocity using a Givens rotation-based eigenvalue decomposition (EVD) instead of the inbuilt AMD Xilinx IP core used previously in \cite{tewari2024reconfigrsp, mansi2022spatialsensing}. Prior works on the hardware implementation of MUSIC employed the Jacobi method for EVD, whereas \cite{MUSIC_HW2} utilized deep learning. Though both offer good performance, they require high complexity \cite{li2022musictcas1,zhang2023HW_EVD_tvt}. Instead, our proposed fixed-point Givens-rotation-based EVD achieves low-latency Doppler velocity estimation with super-resolution capability. Additionally, we propose an adaptive architecture that can switch on-the-fly between MJARP and SARP, suitable for low and high SCNR, respectively. 

{\color{black} We implement reconfigurable RSP on the edge platform- Zynq multiprocessor system-on-chip (MPSoC) via hardware-software co-design (HSCD) between an ARM microprocessor and a field programmable gate array (FPGA). 
\textcolor{black}{The MPSoC is a reliable edge platform offering on-the-fly reconfiguration and has been widely adopted for standalone communication \cite{ tewari2022reconfigofdm, sharma2024dlforce_tcas1, Drozdenko2018hwswzynq} and radar systems \cite{li2022musictcas1, schweizer2021fairy, Lin2024HWspaceborneSAR, zhang2023HW_EVD_tvt}; however, unlike prior works, the proposed work integrates both sensing and communication for ISAC on an edge platform.} 
\textcolor{black}{We compare our proposed ISAC with commercially available millimeter wave radar SoC platforms from Texas Instruments \cite{TI_mmWaveRadarSensors}, Infineon \cite{Infineon_24GHzRadarIoT}, and NXP Semiconductors \cite{NXP_SAF85XXL}. As highlighted in Table~\ref{tab:radar_comparison}, existing 
SoCs are mainly designed for sensing, and their extension to ISAC is not available yet.}
\begin{table}[h!]
    \color{black}
    \centering
    \caption{\textcolor{black}{Comparison of commercially available radar SoC}}
    \label{tab:radar_comparison}
    \begin{adjustbox}{max width=0.5\textwidth}
    \begin{tabular}{lcc}
        \toprule
        \textbf{Parameters} & \textbf{Commercial Radar SoC} & \textbf{Proposed Radar SoC} \\
        \midrule
        \textbf{Application} & Sensing & Sensing and Communication \\
        \textbf{Waveform Type} & FMCW & \textcolor{black}{(phase coded, FMCW, OFDM)} \\
        \textbf{Peak to Side Lobe Level} & $-13.26$ dB & $-96.33$ to $-385.52$ dB (Golay) \\
        \textbf{DAC/ADC Sampling Rate} & $4-45$ MSPS & $1.76$ GSPS \\
        \textbf{Pulse Duration} & High & Low \\
        \textbf{Bandwidth} & Up to $5$ GHz & $1.76$ GHz \\
        \textbf{Range Resolution} & $0.03$ m & $0.085$ m \\
        \textbf{No. of Channels} & Up to 4 & $8 \text{ and } 32$\\
        \textbf{Angular Resolution} & $2.8^\circ\, \text{to}\, 6^\circ$ & $1^\circ$ \\
        \textbf{Reconfigurability} & Parameters & Parameters and Algorithms \\
        \textbf{Fixed Point Implementation} & Float and Integer (16, 32 bits) & Float, Integer and Fixed-point \\
        \bottomrule
    \end{tabular}
     \end{adjustbox}
\end{table}
Our proposed RSP, designed specifically for ISAC and, when implemented on a Zynq MPSoC, provides acceleration up to 5.6 times that of JARP in \cite{tewari2024reconfigrsp} for 3D localization (range, azimuth, and Doppler velocity) of multiple mobile targets. The low-complexity localization (SARP) offers a 51\% reduction in RSP time compared to MJARP and increases ISAC communication throughput by 24\%. The key innovations of the work are:
\begin{enumerate}[wide, labelwidth=!,labelindent=0pt]
  {\color{black}  \item \textbf{Algorithmic Level:} We develop two RSP frameworks for the ISAC system: a low computationally complex, faster sequential processing framework, SARP, and a more accurate but computationally complex and slower \textit{modified} joint processing framework, MJARP. 
    \item \textbf{Architecture Level:} We map both of these algorithms to architectures on Zynq MPSoC and offload range, azimuth, and Doppler estimation for multiple targets to the programmable logic/FPGA, which provides a significant gain in latency. For a multiple-target case, we implement the CLEAN algorithm on the FPGA to remove the contributions of detected targets from the received signal, thereby reducing data communication overhead between the ARM and the FPGA. 
    We develop single-precision floating point (SPFL) and fixed-point hardware IPs for the same and validate the RSP performance under varied SCNR conditions. We create a hardware IP for Doppler estimation via MUSIC with Givens rotation-based EVD. This implementation outperforms the inbuilt AMD Xilinx QRF IP by offering super-resolution capabilities. 
    \item \textbf{System Level:} We develop an efficient FPGA-agnostic reconfigurable architecture that shares the hardware blocks between MJARP and SARP, along with the scheduler to control the hardware blocks. We demonstrate run-time reconfiguration between the SARP-MJARP RSP approaches based on the channel conditions estimated by the SCNR sensing unit to maximize the communication link throughput for multiple targets. In the end, we present system-level ISAC performance for a network with 10 targets. }
\end{enumerate}
\indent The paper is organized as follows. The following section presents the ISAC system model and RSP algorithmic innovations. This is followed by the hardware mapping of the RSP algorithms onto the SoC and the reconfigurable architecture in Section \ref{Sec:RSP_hw}. The functional accuracy and RSP performance analysis are discussed in Section \ref{sec:RSP_PerAnal}, followed by the hardware complexity in Section \ref{Sec:hw_cmpl} and system-level ISAC performance analysis in Section \ref{sec:results}. Section~\ref{Sec:conc} concludes the paper.

\textit{Notation:} Bold lowercase and uppercase letters represent vectors and matrices, respectively, whereas regular characters are used for scalar variables. 
$\mathbf{X}[p,l,m]$ denotes a value of the 3D data cube of $P\times L \times M$ dimensions. 
$\mathbf{X}_m$ refers to the $m^{th}$ matrix slice of size $P \times L$, 
while $\mathbf{X}_{m,l}$ is a length-$P$ vector at indices $m$ and $l$. 
[.] and $(.)$ parentheses denote discrete and continuous-time sequences and signals, respectively. The superscripts $T$ and $*$ denote the transpose and the complex conjugate transpose, respectively. Every $n^{th}$ iteration of variable $\Gamma$ is denoted by $\Gamma^{\{n\}}$. Symbols $\circledast$ and $\odot$ indicate the convolution operation and Hadamard product, respectively. $\mathbb{R}$,
$\mathbb{B}$ and $\mathbb{C}$ correspond to real, binary, and complex data, respectively. 

\section{RSP System Model and Algorithm}
\label{Sec:RSPSysModel}
We consider an ISAC BS transceiver that uses the 802.11ad waveform for both radar and communication, supporting hardware sharing (digital and analog frontends) between these functionalities via time-division multiplexing \cite{sneh2022ieee}. The Doppler-resilient complementary Golay sequences, $\mathbf{G}=[\mathbf{g}_{0},\cdots,\mathbf g_{m}]$ across $m = 1:M$, 802.11ad communication packet preambles/slow time pulses, exhibit excellent autocorrelation properties for range estimation of moving targets \cite{duggal2020doppler}. We transmit the Golay sequences 
at a pulse repetition interval (PRI), $\tau_{PRI}$, with $P=1024$ fast time samples in each pulse 
after passing through the digital-to-analog converter (DAC) with a sampling time of $\Delta\tau$ seconds, as shown in
\par\noindent\small
\begin{align}
\small
\mathbf{z}(\tau) = \sum_{m=0}^
{M-1} \sum_{p=0}^{P-1} \mathbf{G}[m,p]\circledast\delta(\tau-p\Delta\tau-m \tau_{PRI}).
\label{eq:DAC}
\end{align}\normalsize
Then, $\mathbf{z}$ is upconverted to mmWave wavelength, $\lambda$, and radiated through an omnidirectional antenna from the BS transmitter. We consider $T$ point-scatterer targets (or a single extended target with $T$ point scatterers) where each $t^{th}$ target with $\sigma_t$ radar cross-section, is located at a range $r_t$, an azimuth of $\phi_t$, and moving with a Doppler velocity of $v_{t}$ with respect to the ISAC transceiver. 
The radar echoes scattered by the targets fall on an $L$ element uniform linear array (ULA) with inter-element spacing, $d$, at the radar receiver after two-way propagation. After downconversion and digitization, the received signal data cube is 
$\mathbf{X} \in \mathbb{C}^{P \times L \times M}$ - a function of fast time samples ($p$), antenna elements ($l$), and slow time samples ($m$) - 
\par\noindent\small
\begin{align}
\label{eq:RxSig}
\small
\begin{split}
\mathbf{X}\left[p,l,m\right]= \sum_{{t=1}}^T \left[\frac{A}{r_{t}^2}\sqrt{\frac{\sigma_t\ricK}{\ricK+1}} + \rho_{t}\sqrt{\frac{1}{\ricK+1}}\right]\\ \mathbf{G}\left[m,p-\frac{2r_{t}}{c\Delta\tau}\right]e^{j\frac{2\pi}{\lambda}ld\sin\phi_t}e^{j \frac{4\pi}{\lambda} v_{t} m \tau_{PRI}}
+ \eta.\\ 
\end{split}
\end{align}\normalsize
Here, $A$ represents the Friis' free space path-loss, incorporating the transmit power, and radar antenna gains, and $\eta$ denotes the complex additive white Gaussian noise (AWGN) 
The wireless channel is represented with a generalized Rician model with Rician factor, $\ricK$. The first and second components of \eqref{eq:RxSig} show the strength of the line-of-sight (LOS) and non-line-of-sight (NLOS) returns, respectively. High values of $\ricK$ indicate a dominant LOS path, 
while lower values of $\ricK$ represent multipath dominating the propagation with strength, $\rho_t$, modeled by a complex Gaussian random distribution.

$\mathbf{X}$ is first processed along $p$ and $l$ dimensions to localize targets in range-azimuth (RA). We briefly recapitulate the joint azimuth-range processing (JARP) framework for RA localization, as discussed in \cite{tewari2024reconfigrsp}. Then, we present the algorithmic modifications and computational gains in the proposed modified JARP (MJARP) and sequential azimuth range processing (SARP) algorithms. Once multiple targets are detected post-RA localization, we process the slow-time cells corresponding to each target to estimate Doppler velocity. 
\subsection{MJARP vs JARP}
The RA processing is carried out on 
$\mathbf{X}_{m}$, corresponding to each $m^{th}$ slow-time pulse.
\subsubsection{JARP}
Here, we perform FFT on the $P$ fast time samples of each $l^{th}$ antenna channel of $\mathbf{X}_{m}$ to obtain the frequency domain representation, $\mathbf{\mathbf{\tilde{X}}}_{m} \in \mathbb{C}^{P \times L}$. 
\par\noindent\small
\begin{align}
\label{eq:2D_fft}
\small
\begin{split}
\mathbf{\tilde{X}}_{m,l}=\fftp\{\mathbf{X}_{m,l}\}.
\end{split}
\end{align}\normalsize
Then, the signal undergoes digital beamforming (DBF) across antennas and matched filtering (MF) across fast-time samples through cross-correlation between the transmitted and received radar signals in the frequency domain. 
We consider the target's angular search space $\Omega$, discretized into $I$ search angles for DBF, with a step size of $\Delta \phi$, $\phi=[0,\Delta\phi,2\Delta\phi,\cdots, (I-1)\Delta\phi]$. The RA processing is carried out for each $m^{th}$ 2D packet to obtain the RA ambiguity plot for the $m^{th}$ pulse, $\mathbf{\Gamma}_m \in \mathbb{C}^{I \times P}$, as shown in, 
\par\noindent\small
\begin{align}
\label{eq:2D_mf}
\begin{split}
\small
\mathbf{\Gamma}_{m}\left[\phi,r\right]=\ifftp(\mathbf{\tilde{X}}_{m}\mathbf{W}_{\phi}^T\odot\mathbf{\tilde{g}}^\ast_{m}).
\end{split}
\end{align}\normalsize
Here, ${\mathbf{\Tilde{g}}_{m}}=\fftp(\mathbf{g}_{m}) \in \mathbb{C}^{P\times1}$ is the frequency domain response of the transmitted sequence and $\mathbf{W} \in \mathbb{C}^{I \times L}$ 
is the antenna weight matrix comprising $\mathbf{W}_{\phi}=\left[1,\cdots,e^{-j\frac{2\pi}{\lambda}(L-1)d\sin\phi} \right]$ for every $\phi$. 

When there are multiple targets, the weaker targets may be masked by the RA sidelobes of the stronger targets in $\mathbf{\Gamma}_0$. So, we perform the iterative CLEAN algorithm to detect multiple weak targets in the RA ambiguity plot of the first packet, $\mathbf{\Gamma}_0$. In each $n^{th}$ iteration of CLEAN, we estimate the range and azimuth corresponding to the peak of $\mathbf{\Gamma}_0^{\{n\}}$: $<\hat{\phi}^{\{n\}},\hat{r}^{\{n\}}>=\argmax{\phi,r}|\mathbf{\Gamma}_0^{\{n\}}|$. We use these estimates to simulate the time-domain dummy target returns as shown in  
\par\noindent\small
\begin{align}
\mathbf{S}^{\{n\}}[p,l] = \hat{a}^{\{n\}}\mathbf{G}\left[m=0,p-\frac{2\hat{r}^{\{n\}}}{c\Delta\tau}\right]e^{j\frac{2\pi}{\lambda}ld\sin{\hat{\phi}^{\{n\}}}},
\end{align}\normalsize
where $\hat{a}^{\{n\}} = \max|\mathbf{\Gamma}_0^{\{n\}}|$.
We then generate the point spread response (PSR) of this signal in the RA domain by following the steps in \eqref{eq:2D_fft}, \eqref{eq:2D_mf} to obtain $\mathbf{\breve{\Gamma}}_{0}^{\{n\}}\in \mathbb{C}^{I \times P}$. This PSR is subtracted from the RA plot to obtain the residue plot $\mathbf{\Gamma}_0^{{\{n+1\}}}$. This process is repeated till the detected peak amplitude, $|\hat{a}^{\{n\}}|$, falls below the radar's minimum detectable signal strength threshold.
In each iteration, the strongest target's contributions (main and sidelobes) are removed from the RA plots so that weaker targets may be detected more easily. Once, all the targets are localized in range and azimuth from the first packet, RA processing is conducted on the remaining $(M-1)$ packets to create the slow time vector, $\mathbf{\bm{\zeta}}^{\{n\}} = \left[\mathbf{\Gamma}_0[\hat{\phi}^{\{n\}},\hat{r}^{\{n\}}],\mathbf{\Gamma}_1[\hat{\phi}^{\{n\}},\hat{r}^{\{n\}}], \cdots, \mathbf{\Gamma}_{(M-1)}[\hat{\phi}^{\{n\}},\hat{r}^{\{n\}}] \right]$ by selecting only the elements corresponding to the $n^{th}$ target localization coordinates $[\hat{\phi}^{\{n\}},\hat{r}^{\{n\}}]$ from every $m^{th}$ pulse.
\subsubsection{Modified Joint Azimuth-Range Processing (MJARP)}
In this work, we optimize the computational effort required to process subsequent packets ($m > 0$) by leveraging precomputed estimates from the first pulse. Here, DBF is only performed for the angle $\hat{\phi}^{\{n\}}$, and the IFFT
operation is replaced with a single $P$ element multiply and accumulate (MAC) operation with the Fourier vector corresponding to $\hat{r}^{\{n\}}$. For each $n^{th}$ target, the first slow time sample is extracted as $\mathbf{\bm{\zeta}}^{\{n\}}[0]=\mathbf{\Gamma}_0[\hat{\phi}^{\{n\}},\hat{r}^{\{n\}}]$, and the subsequent $(M-1)$ samples are derived as follows, 
\par\noindent\small
\begin{align}
\begin{split}
\label{eq:V4}
\small
\mathbf{\bm{\zeta}}^{\{n\}}\left[m\right]= \sum_{{p=0}}^{P-1}\left[\mathbf{\Tilde{X}}_{m,p}\mathbf{W}_{\hat{\phi}^{\{n\}}}{\mathbf{\Tilde{g}}}^\ast_{m}[p]e^{-j2\pi \frac{2 {\hat{r}^{\{n\}}} p}{c \Delta\tau}}
\right]
\end{split}
\end{align}\normalsize
Figures \ref{fig:rsp_cmp}(a) and (c) show the RA processing for $M$ packets in MJARP.
\begin{figure}[!ht]
    \centering
    \vspace{-0.2cm}
    \includegraphics[scale = 0.28]{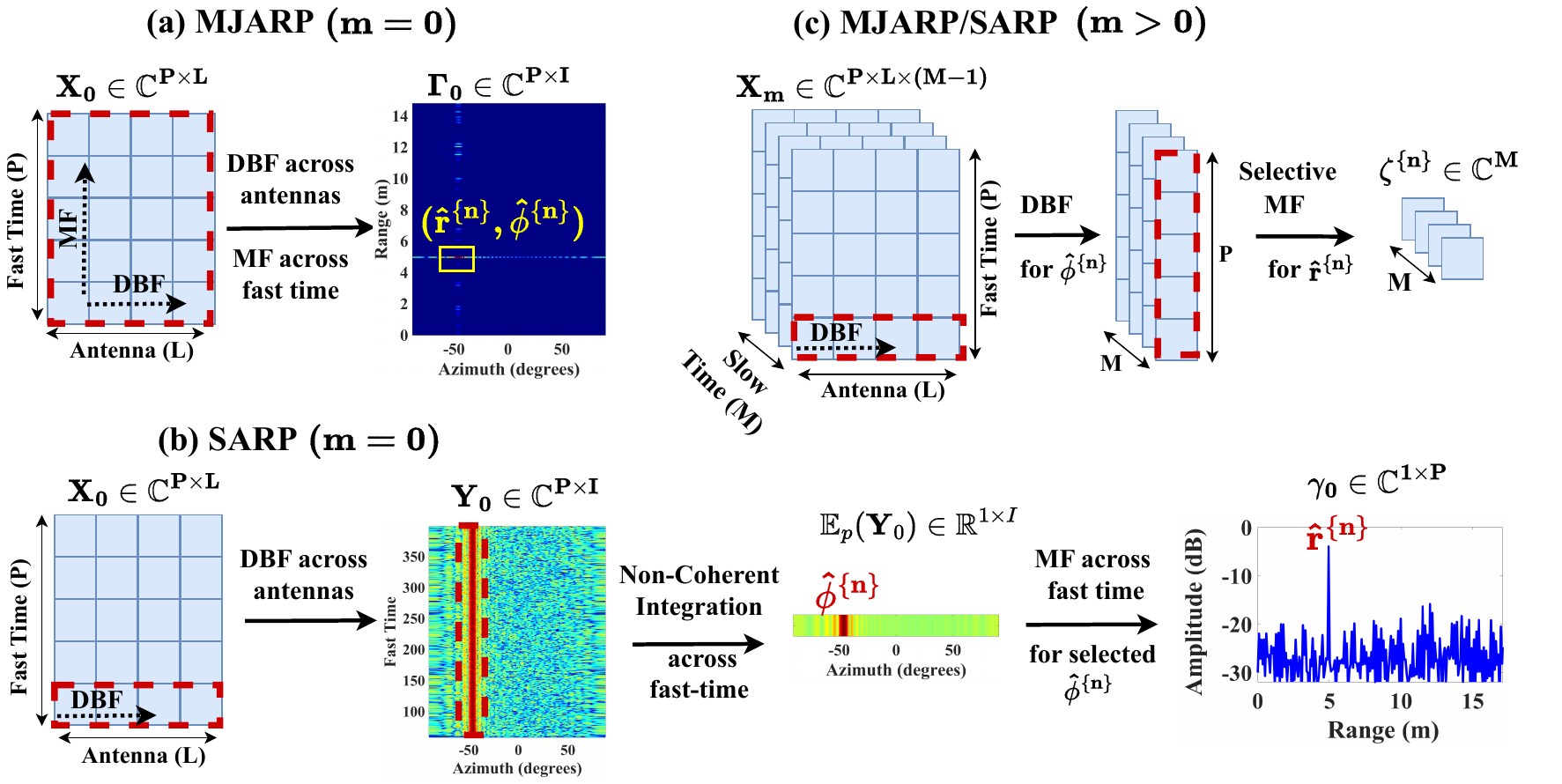}
    \vspace{-0.2cm}
    \caption{\small DBF and MF for RA localization of $n^{th}$ target with (a) MJARP for first $(m=0)$ packet, (b) SARP for first $(m=0)$ packet, (c) MJARP and SARP for next $(M-1)$ packets}
    \label{fig:rsp_cmp}
\end{figure}
\subsection{Sequential Azimuth-Range Processing (SARP)}
In MJARP, the range and azimuth of each target are jointly estimated from the 2D raw data square. We propose an alternative low-complexity architecture based on SARP as shown in Figure \ref{fig:rsp_cmp}. (b). First, DBF is performed on the first packet, $\mathbf{X}_0$, across all search angles to obtain
\par\noindent\small
\begin{align}
\label{eq:1D_dbf}
\begin{split}
\small
\mathbf{Y}_{0}\left[\phi,p\right]=\mathbf{X}_{0,p}\mathbf{W}_{\phi}^T
\end{split}
\end{align}\normalsize
We improve the azimuth accuracy by non-coherently integrating across the fast time samples for each $\phi$, to obtain,
\par\noindent\small
\begin{align}
\label{eq:dbf_avg}
\expct(\mathbf{Y}_{0}^{\{n\}})=\frac{1}{P}\sum_{p=0}^
{P-1} |\mathbf{Y}_{0}^{\{n\}}\left[\phi,p\right]|.
\end{align}\normalsize
Then we estimate the angular position corresponding to the peak 
on $\hat{\phi}^{\{n\}} = \argmax{\phi}\expct(\mathbf{Y}_{0}^{\{n\}})$.
We select the vector $\mathbf{Y}_{0,{\hat{\phi}^{\{n\}}}}$ for range processing through MF, which involves its cross-correlation with the transmitted sequence, $\mathbf{g}_0$, and can be carried out in time-domain by
\par\noindent\small
\begin{align}
\label{eq:1D_mftd}
\bm{\gamma}\left[r\right]= \sum_{p=0}^{P-1}\mathbf{Y}_{0,\hat{\phi}^{\{n\}}}[p-r]\mathbf{g}_0^*[p].
\end{align}\normalsize
Alternatively, we can consider frequency domain MF (FD-MF) approach.
First, we FFT the selected vector, $\Tilde{\mathbf{y}}=\fftp([\mathbf{Y}_{0,\hat{\phi}^{\{n\}}}])$ followed by element-wise multiplication with FFT of first transmit sequence, $\tilde{\mathbf{g}}_0$, and IFFT. 
\par\noindent\small
\begin{align}
\label{eq:1D_mffd}
\begin{split}
\small
\bm{\gamma}\left[r\right]=\ifftp(\mathbf{\tilde{y}}\odot \mathbf{\Tilde{g}}^\ast_{0}) 
\end{split}
\end{align}\normalsize
Now, peak search is conducted over $\bm{\gamma} \in \mathbb{C}^{P \times 1}$ to obtain the range estimate, $<\hat{r}^{\{n\}}>=\argmax{r}\bm{\gamma}$.
Once the strongest target is localized, CLEAN is performed on the first packet to detect weak targets. The detection parameters of the $n^{th}$ target, $<\hat{a}^{\{n\}},\hat{r}^{\{n\}},\hat{\phi}^{\{n\}}>$, are used to generate the PSR. Unlike the MJARP processing, PSR in SARP is the image after DBF, as shown in, 
\par\noindent\small
\begin{align}
\label{eq:1D_psr}
\breve{\mathbf{Y}}_0^{\{n\}}\left[\phi,p\right]=\mathbf{S}_{0,p}^{\{n\}}\mathbf{W}_{\phi}^T 
\end{align}\normalsize
\normalsize
Here, $\mathbf{S}_{0,p}^{\{n\}}\in \mathbb{C}^{P \times L}$, is generated as shown in Section \ref{Sec:RSPSysModel}.A.1. Peak search on the residue, $\mathbf{{Y}}_{0}^{\{n+1\}}=\mathbf{{Y}}_{0}^{\{n\}}-\mathbf{\breve{Y}}_{0}^{\{n\}}$, provides the localization estimates of the $(n+1)^{th}$ target. 
\subsection{Computational Complexity Comparison}
Table. \ref{tab:MAC_cmp} lists the number of computational operations required for the RA processing for a single target across $M$ slow-time packets with JARP, SARP, and MJARP RSP architectures. The number of computations in the first $m=0$ packet is identical for JARP and MJARP, whereas SARP requires fewer computations. 
For the remaining $m=1\cdots M-1$ packets, the RA processing is identical in the case of SARP and MJARP. Here, the selective MF in SARP/MJARP results in a significant reduction in complex MAC (CMAC) and complex multipliers (CM) operations as compared to JARP. Thus, for $M$ packets, SARP RSP has the least number of CMAC/CM operations, offering the lowest latency, followed by MJARP and then JARP. 
MJARP, with DBF across the $L$ antenna channels and cross-correlation across $P$ fast-time samples, offers a coherent processing gain of $\frac{PL}{2}$. Hence, even under low  SCNR, the signal strength post-MJARP is significantly amplified, leading to better estimation than SARP.

\begin{table}[h]
\caption{\small Complexity comparison between JARP, MJARP, and SARP}
\label{tab:MAC_cmp}
\huge
\renewcommand{\arraystretch}{1.05}
\resizebox{\linewidth}{!}{
\centering
\renewcommand{\arraystretch}{1.2}
\begin{tabular}{|c|c|ccc|c|}
\hline
\multirow{2}{*}{\textbf{\begin{tabular}[c]{@{}c@{}}Slow Time\\Packets\end{tabular}}} &
  \multirow{2}{*}{\textbf{RSP}} &
  \multicolumn{3}{c|}{\textbf{CMAC}} &
  \multirow{2}{*}{\textbf{CM}} \\ \cline{3-5}
 &                & \multicolumn{1}{c|}{\textbf{DBF}} & \multicolumn{1}{c|}{\textbf{FFT}} & \textbf{IFFT} &    \\ \hline
\multirow{2}{*}{\textbf{$m=0$}} &
  \begin{tabular}[c]{@{}c@{}}\textbf{JARP/}\\\textbf{MJARP}\end{tabular} &
  \multicolumn{1}{c|}{$PLI$} &
  \multicolumn{1}{c|}{$\frac{PL}{2}\log_{2}P$} &
  $\frac{PI}{2}\log_{2}P$ &
  $PI$ \\ \cline{2-6} 
 & \textbf{SARP}  & \multicolumn{1}{c|}{$PLI$}          & \multicolumn{1}{c|}{$\frac{P}{2}\log_{2}P$}     & $\frac{P}{2}\log_{2}P$      & $P$  \\ \hline
 \multirow{3}{*}{\begin{tabular}[c]{@{}c@{}}\textbf{$m=1\cdots$}\\(M-1)\end{tabular}} &
  \textbf{JARP} &
  \multicolumn{1}{c|}{$PLI(M-1)$} &
  \multicolumn{1}{c|}{$\frac{PL(M-1)}{2}\log_{2}P$} &
  $\frac{PI(M-1)}{2}\log_{2}P$ &
  $PI(M-1)$ \\ \cline{2-6} 
 &   \multicolumn{1}{c|}{\multirow{2}{*}{\begin{tabular}[c]{@{}c@{}}\textbf{MJARP/}\\\textbf{SARP}\end{tabular}}} & \multicolumn{1}{c|}{\multirow{2}{*}{$PL(M-1)$}} & \multicolumn{1}{c|}{\multirow{2}{*}{$\frac{PL(M-1)}{2}\log_{2}P$}}  & \multirow{2}{*}{$P(M-1)$}  & \multirow{2}{*}{$P(M-1)$} \\ & \multicolumn{1}{c|}{} & \multicolumn{1}{c|}{} & \multicolumn{1}{c|}{} & \multicolumn{1}{c|}{}  & \multicolumn{1}{c|}{}\\ \hline
\end{tabular}}
\end{table}

\subsection{Doppler Velocity Estimation}
After CLEAN, once all potential targets are localized in RA, Doppler velocity estimation is performed sequentially on each detection to distinguish mobile targets from static clutter. The slow time vectors corresponding to each target, $\mathbf{\bm{\zeta}}^{\{n\}}\in \mathbb{C}^{M \times 1}$, are sent to the MUSIC algorithm for Doppler velocity estimation. 
MUSIC comprises three stages- spatial smoothing (SS) for averaged covariance generation, EVD for separating out the signal and noise subspaces, and MUSIC spectrum generation (MSG), as discussed in \cite{tewari2024reconfigrsp}. We implement EVD using QR factorization with the Givens Rotation method. it is performed on the averaged covariance matrix, $\mathbf{U}\in \mathbb{C}^{K \times K}$, where $K$ is the resultant slow time snapshot length post-averaging in the SS stage, such that $K<M$. In QR factorization, $\mathbf{U}$ is decomposed into an orthogonal matrix, $\mathbf{Q} \in \mathbb{C}^{K \times K}$, and an upper triangular matrix, $\mathbf{R} \in \mathbb{C}^{K \times K}$, as $\mathbf{U} =\mathbf{Q}\mathbf{R}$. The Givens rotation method, its hardware, and advantages over state-of-the-art are detailed in Section \ref{Sec:RSP_hw}. E.

\section {Mapping of Algorithms to Architecture}
\label{Sec:RSP_hw}
 We implement the reconfigurable RSP on Zynq MPSoC, as shown in Figure \ref{fig:rsp_mpsoc}. The various radar tasks are partitioned between processing system (PS) and programmable logic (PL) via HSCD for an efficient high-speed implementation. The target modeling, channel modeling, target-reflected received packet generation, and results display are carried out in PS. All the RSP tasks, such as MF, DBF, MUSIC, peak search, and CLEAN are offloaded to the RSP accelerator in PL. The PS transmits the received data packets to PL via Direct Memory Access (DMA) and also provides the RSP configuration information via the AXI-Lite port. Based on the configuration setting, which includes the angular precision $\Delta \phi$, number of slow-time packets $M$, and RSP algorithm- SARP or MJARP, the PL accelerator independently schedules and implements the operations. The PL architecture is made reconfigurable and offers run-time switching between the different $M$, $\Delta \phi$, and RSP algorithms.
\begin{figure}[!ht]
    \centering
    \vspace{-0.2cm}
    \includegraphics[scale = 0.42]{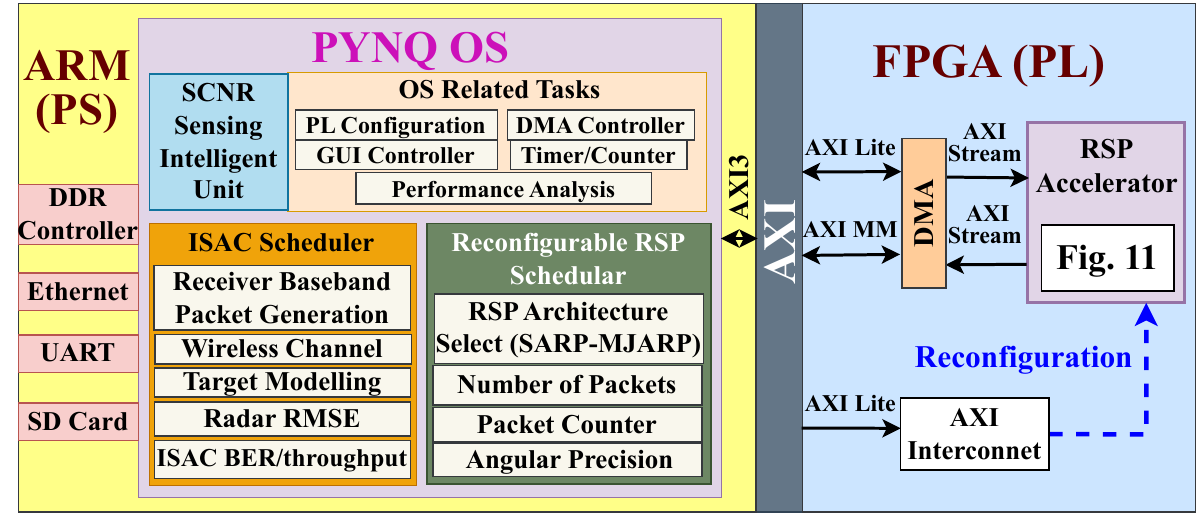}
    \vspace{-0.2cm}
    \caption{\small Reconfigurable RSP on Zynq MPSoC via HSCD.}
    \label{fig:rsp_mpsoc}
\end{figure}
This section discusses mapping the various RSP blocks described in Section \ref{Sec:RSPSysModel} on a reconfigurable architecture. First, MF and DBF are carried out on the first $(m=0)$ 2D pulse, $\mathbf{X}_{0}$, with SARP or MJARP approaches, followed by peak search. Then, CLEAN is performed on the first received pulse to identify multiple targets in the RA space. Once localized, we perform selective RA processing only along the detections $<\hat{\phi}^{\{n\}},\hat{r^{\{n\}}}>$. This is followed by Doppler processing for each detected target. 
\subsection{Range-Azimuth Localization on first packet with SARP}
\label{Sec:1DRSP_arch}
As shown in equation \eqref{eq:1D_dbf}, the first step is the DBF of $\mathbf{X}_{0}$. Figure \ref{fig:dbf} shows the DBF architecture. Here, $\mathbf{X}_{0}$ is stored in BRAM $A$ via DMA by PS. BRAM $B$ contains the $L$-element beamforming weight vectors for each of the $I$ search angles, $W\in\mathbb{C}^{I\times L}$. There are two counters for tracking the search angles and fast time samples, initialized in stage $C0-C1$. In stage $C2$, the $L$ element rows corresponding to the fast-time index and angular search index are read from BRAM $A$ and $B$, respectively, and sent to the L-input MAC unit. In stage $C3$, element-wise multiplication occurs between the input and the weight vector via $L$ parallel CM. The result is accumulated using complex adder (CA) to form a single complex value written in BRAM $C$ in stage $C4$. Since this process is performed $P \times I$ times (for each fast time sample and every search angle), the parallelism in the $L$-input MAC unit results in a significant speed-up. Also, since the DBF is carried out sequentially for every $\phi$, the architecture enables on-the-fly reconfigurability for different angular precisions, $\Delta \phi$, by appropriately configuring the read operation from BRAM $B$. 
\begin{figure}[!t]
    \centering
    \vspace{-0.2cm}
    \includegraphics[scale = 0.7]{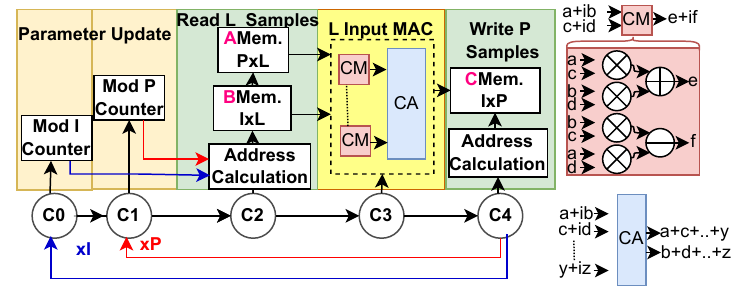}
    \vspace{-0.2cm}
    \caption{\small Hardware architecture for SARP DBF.}
    \label{fig:dbf}
     \vspace{-0.2cm}
\end{figure}
\begin{figure}[!b]
    \centering
    \vspace{-0.25cm}
    \includegraphics[scale = 0.625]{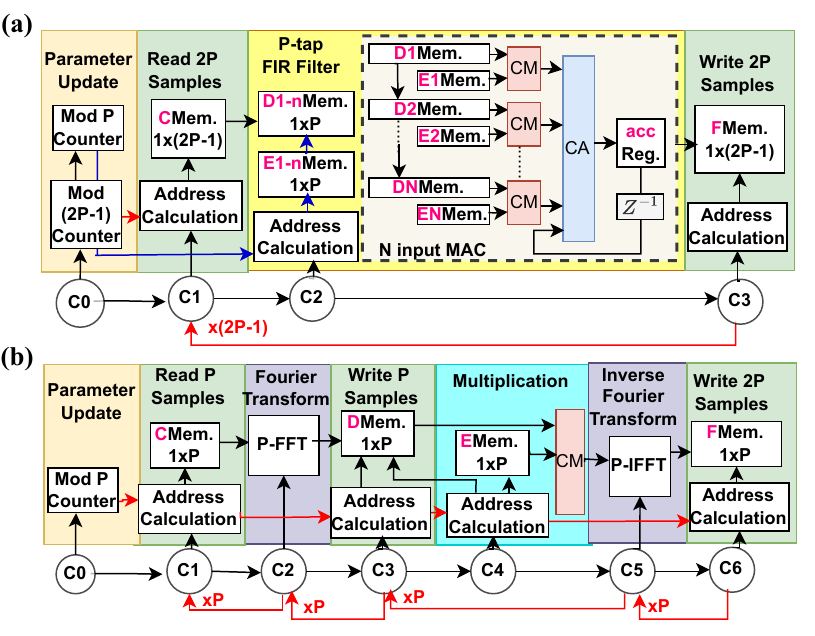}
    \vspace{-0.25cm}
    \caption{\small Hardware architecture for SARP (a)TD-MF, (b)FD-MF.}
    \label{fig:rsp_1d}
     \vspace{-0.25cm}
\end{figure}
\begin{figure*}[!b]
    \centering
    \vspace{-0.2cm}
    \includegraphics[scale = 0.85]{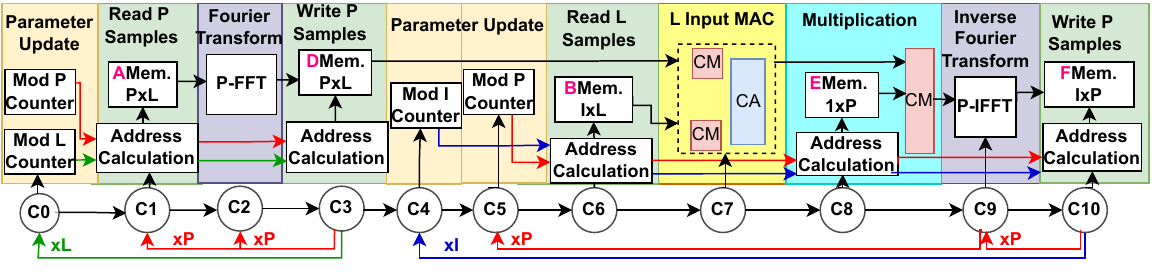}
    \vspace{-0.2cm}
    \caption{\small Hardware architecture for MJARP MF and DBF.}
    \label{fig:MF_2D}
\end{figure*}
After the peak search, the vector $\mathbf{Y}_{0,\hat{\phi}^{\{n\}}}$, as mentioned in Section \ref{Sec:RSPSysModel}.B, is selected for further processing with MF. We explore two architectures- time domain and frequency domain MF as explained below. 
\subsubsection{Time Domain Matched Filtering}
We use the finite impulse response (FIR) filter for implementing the time-domain cross-correlation between $\mathbf{Y}_{0,\hat{\phi}^{\{n\}}}$ and the transmitted sequence, $\mathbf{g}_0$, as shown in equation \eqref{eq:1D_mftd}. The architecture is shown in Figure~\ref{fig:rsp_1d} (a). The vector $\mathbf{Y}_{0,\hat{\phi}^{\{n\}}} \in \mathbb{C}^{P \times 1}$ is stored in BRAM $C$ with the last $P-1$ elements initialized to zeroes. The P-tap FIR comprises $\mathbf{g}_{0} \in \mathbb{C}^{P\times1}$ as the $P$ filter weights stored in BRAM $E$, a $P$-length first-in-first-out (FIFO) shift register (initialized with zeroes), and a MAC unit. The counters are initialized in stage $C0$. In $C1$, a new complex sample is read from BRAM $C$ and sent to the FIR. $C2$ is the FIR filtering stage where the sample enters the FIFO shift register, and $P$ element-wise complex MAC operations take place between the shift register elements and the filter weights. The accumulated value is written in BRAM $F$ in stage $C3$. This process is carried out for each of the $2P-1$ input samples. 
We also explore a parallelized architecture by partitioning the shift register and weights into $N$ BRAMs and using $N$ parallel CM units. 
\subsubsection{Frequency Domain Matched Filtering}
The vector $\mathbf{Y}_{0,\hat{\phi}^{\{n\}}}$ is stored in BRAM $C$. As shown in Figure \ref{fig:rsp_1d} (b). We use the built-in AMD Xilinx FFT IP for the FFT and IFFT. During stage $C1$, the $P$ elements are read from BRAM $C$ and sent to the FFT core. After the $P$-point FFT operation in $C2$, the output is written back in BRAM $D$ in stage $C3$. The FFT of the transmit sequence, $\mathbf{\tilde{g}^\ast_{0}}\in \mathbb{C}^{P \times 1}$ is stored in BRAM $E$. During stage $C4$, based on the fast time index, the corresponding samples are read from BRAM D and E, multiplied, and sent to the IFFT. 
After $P$-point IFFT in stage $C5$, the $P$ samples are sequentially written in BRAM $F$. Unlike TD-MF, parallelizing the CM operations is not useful for FD-MF, as FFT and IFFT work on stream data.
\subsection{Range-Azimuth Localization on first packet with MJARP}
\label{Sec:2DRSP}
The MJARP is performed on $\mathbf{X}_{0}\in \mathbb{C}^{P \times L}$, as shown in equations \eqref{eq:2D_fft} and \eqref{eq:2D_mf}. $\mathbf{X}_{0}$ is stored in BRAM $A$, as shown in Figure \ref{fig:MF_2D}. $P$-point FFT is performed sequentially for each of the $L$ antenna channels in stages $C0-C3$ and $\mathbf{\tilde{X}}_{0}\in \mathbb{C}^{P \times L}$ is stored in BRAM $D$. After this, the DBF and MF are pipelined for each fast-time sample in stages $C4-C8$. The beamforming weights and the correlation sequence are stored in BRAM $B$ and $E$, respectively. In stage $C6$, $L$ element input and weight vectors are selected from the BRAMs and undergo parallelized MAC operations in stage $C7$ for DBF. The output sample is multiplied by the selected correlation sample in stage $C8$ and sent to the IFFT unit. After $P$-point IFFT in $C9$, the samples are written to BRAM $F$ in $C10$. Stages $C4-C10$ are carried out sequentially for each of the $I$ search angles. Therefore, unlike the SARP approach with only a single $P$-point FFT, $P$-element CM, and $P$-point IFFT, the MJARP approach has $L$ times $P$-point FFTs and $I$ times $P$-element CM, and $P$-point IFFTs (as mentioned in Table~\ref{tab:MAC_cmp}), resulting in higher execution time. 
In MJARP, we prefer FD-MF as SARP analysis in Section VI.A indicates that FD-MF is faster.
\subsection{Extension to Multiple Targets (SARP/MJARP)}
The SARP/MJARP RA processing on the first packet is followed by peak search to determine the first target, $T_1$, and CLEAN, which involves PSR generation based on localized estimates of $T_1$ and its complex subtraction from the previous image to determine $T_2$. This process is repeated for the rest of the targets. Since a reconfigurable architecture that switches between SARP and MJARP is preferred, the CLEAN and peak search units are designed such that they can be used in both architectures. The CLEAN for both SARP and MJARP RSP is performed on the output of DBF, $\mathbf{Y}_0^{\{n\}} \in \mathbb{C}^{I \times P}$, as shown in Section \ref{Sec:RSPSysModel}.B, for enabling the run-time switch between SARP and MJARP at any point until termination. 

Figure \ref{fig:clean} illustrates the complex subtraction for CLEAN for both SARP and MJARP, followed by non-coherent fast time integration and azimuth peak search, only in the case of SARP. The complex samples of $\mathbf{Y}_0^{\{n\}}$ are streamed into the CLEAN block one by one and stored in register $in\_cval$. Based on the selected RSP algorithm (SARP/MJARP) and the CLEAN iteration, indicated by the $RSP$ and $CL$ flags, respectively, different operations are scheduled. The $CL$ flag is initialized to zero at the beginning. For $T_1$, since $CL=0$, the control arrives at $C3$, bypassing stage $C1$. Here, the samples are written in BRAM $F$. Depending upon the $RSP$ flag, a simultaneous fast time averaging is performed in the case of SARP,  as discussed in equation \eqref{eq:dbf_avg}, whereas in the case of MJARP, these samples are directly passed ahead for MF. In SARP, the streaming samples are accumulated based on the integration factor (IF), which is the number of fast time samples to be integrated. 
After $P$ iterations, the register $R17$ is updated with the averaged fast time value for the selected search angle. The stage $C4$ is only executed in the case of SARP, where peak search across azimuth takes place. Here, the accumulated value in $R17$ for the current search angle is compared with the existing averaged peak in $R2$. After $I$ iterations, the peak azimuth index is updated in $R2$ for SARP. Also, matrix $\mathbf{Y}_0^{\{n\}}$ is written in BRAM $F$, and the $CL$ flag is set to one now, for SARP and MJARP both. This is followed by MF, which is performed on fast time samples corresponding to all search angles in the case of MJARP, but only for the selected peak azimuth in the case of SARP. 
\begin{figure}[!h]
    \centering
    \vspace{-0.2cm}
\includegraphics[scale = 0.58]{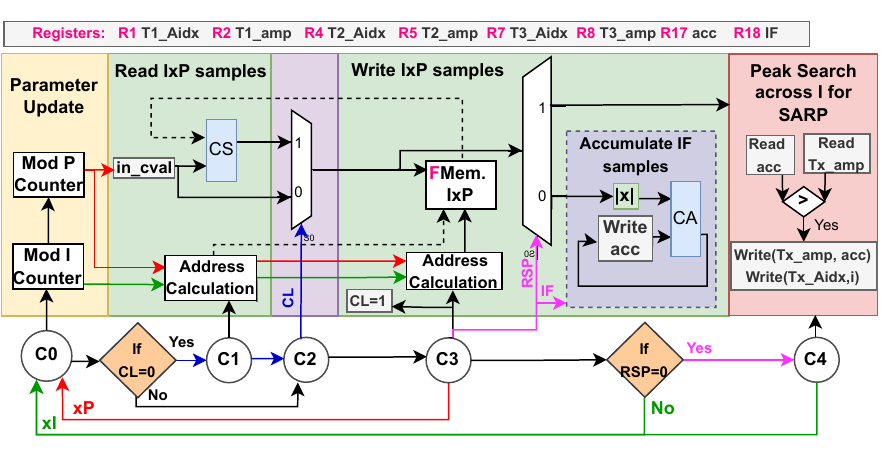}
\vspace{-0.2cm}
    \caption{\small Architecture for CLEAN, and azimuth peak search for SARP. }
    \label{fig:clean}
\end{figure}
Figure~\ref{fig:ps} shows the peak search operations after MF for SARP and MJARP. The peak search is split across the $P$ and $I$ dimensions and calculated in separate units. Here, in stage $C1$, $P$ fast time samples are streamed one by one for fast time peak search. In case of SARP, these samples correspond to $\bm{\gamma} \in \mathbb{C}^{P \times 1}$. After $P$ iterations, the peak range index and amplitude of $T_1$ are written in $R15$ and $R16$ respectively, and this is followed by PSR generation in $C3$. In MJARP, the fast time peak search across each $P$-element column of $\mathbf{\Gamma}_0\in \mathbb{C}^{I \times P}$ takes place in $C1$, followed by peak search across search angles in $C2$. After $P\times I$ iterations, registers $R0-R2$ are updated with $T_1$ peak and this is followed by PSR in $C3$.   
\begin{figure}[!t]
    \centering
    \vspace{-0.2cm}
\includegraphics[scale = 0.75]{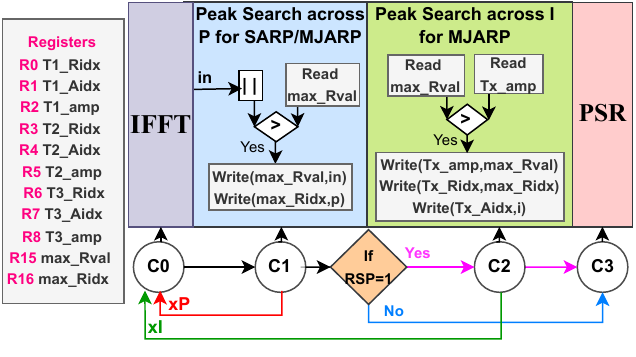}
\vspace{-0.2cm}
    \caption{\small Architecture of peak search for target localization with SARP and MJARP. }
    \label{fig:ps}
\end{figure}

The PSR generation based on the $T_1$ detection parameters, as described in Section \ref{Sec:RSPSysModel}.A.1. is shown in Figure.~\ref{fig:psr}. BRAM $G$ contains the time domain transmit Golay sequence for the first packet, $\mathbf{g_0}\in \mathbf{C}^{\frac{P}{2}\times 1}$. In stages, $C1-C2$, the transmit sequence is read from BRAM $G$, multiplied with the amplitude of $T_1$ stored in register $R2$, and written in BRAM $H$. BRAM $H$ is initialized to zero, and the samples are written at an offset based on the range index value of $T_1$ stored in register $R0$, to account for the time delay due to range. In $C3-C5$, the delayed sequence samples are read one by one, and for each sample, $L$ copies are produced after multiplication with the complex conjugate of the beamforming weight vector read from BRAM $B$ corresponding to the $T_1$ azimuth value read from register $R1$. These $L$ samples are written in BRAM $A$, which is reused again during CLEAN, in stage $C5$. After $P$ iterations, the delayed and phase-shifted matrix for $T_1$, $\mathbf{S}_{0}^{\{n\}} \in \mathbb{C}^{P \times L}$, is stored in BRAM $A$. $\mathbf{S}_{0}^{\{n\}}$ is now processed through DBF for obtaining the PSR, $\mathbf{\breve{Y}}_0^{\{n\}} \in \mathbb{C}^{I \times P}$. 
\begin{figure}[!t]
    \centering
    \vspace{-0.2cm}
\includegraphics[scale = 0.64]{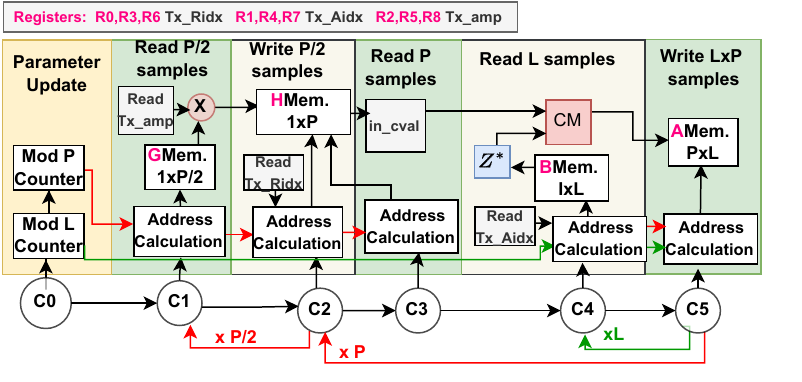}
\vspace{-0.2cm}
    \caption{\small Architecture for PSR generation for CLEAN.}
    \label{fig:psr}
    \vspace{-0.2cm}
\end{figure}

Now, $\mathbf{\breve{Y}}_0^{\{n\}}$ is streamed into the CLEAN block shown in Figure~\ref{fig:clean}. Since the CLEAN flag was set in the previous iteration, the control now enters stage $C1$, where the complex subtraction takes place for the removal of the PSR of $T_1$ from the original image $\mathbf{Y}_{0}^{\{n\}}$. Here, each complex sample of $\mathbf{\breve{Y}}_0^{\{n\}}$ (for the fast-time index $p$ and search angle index $i$) is subtracted from the corresponding sample of $\mathbf{Y}_{0}^{\{n\}}$ read from BRAM $F$ and is rewritten in BRAM $F$ at the same location in stage $C3$. Thus, after $P \times I$ iterations, BRAM $F$ is updated with $\mathbf{Y}_0^{\{n+1\}}$. Also, the residue samples obtained after complex subtractions are simultaneously sent for MF in case of MJARP, or fast time integration and azimuth peak search for SARP. After MF, $T_1$ peak amplitude and indices are updated in registers $R3-R5$. This is followed by the PSR generation and complex subtraction for the localization of $T_3$. 
\label{Sec:clean}
\vspace{-0.2cm}
\subsection{Multi-Packet Range-Azimuth Localization (SARP/MJARP)}
\label{Sec:sel_MF}
After CLEAN on the first packet, all targets are localized, and the respective peak values and indices are stored in the registers $R0-R8$. We conduct only selective RA processing for the remaining packets, as described in equation \eqref{eq:V4}, and depicted in Figure \ref{fig:sel_mf}. The new packet, $\mathbf{X}_{m} \in \mathbb{C}^{P\times L}$ is written in BRAM $A$ and undergoes FFT in stage $C1$. This is followed by DBF, MF, and slow-time vector creation for each target. In stage $C2$, DBF is performed only for the detected azimuth index of the $n^{th}$ target, read from the register, and not for all $I$ search angles. Then, MF takes place in $C4$. Unlike the complete RA processing discussed in Section \ref{Sec:2DRSP}, stages $C2-C5$ are executed just once instead of $I$ times for each sample, thus significantly saving the number of computations. To further reduce the computations, we perform only range-index-specific IFFT 
In stage $C5$, we generate the $P$-element Fourier vector based on the $n^{th}$ target's range index, since storing a hard-coded $P\times P$ Fourier matrix is not feasible due to memory constraints on the FPGA. Each complex sample released from stage $C4$ is multiplied by the corresponding Fourier sample and accumulated in the MAC unit in stage $C5$. This process is repeated for all the $P$ fast-time samples, and the final output of the complex MAC unit is normalized by $P$ and written in BRAM $I$. 
After $M-1$ iterations, the BRAM $I$ is updated with each target's slow-time vector $\bm{\zeta}^{\{n\}} \in \mathbb{C}^{M\times 1}$, ready for Doppler processing.

\begin{figure}[!t]
    \centering
\includegraphics[scale = 0.52]{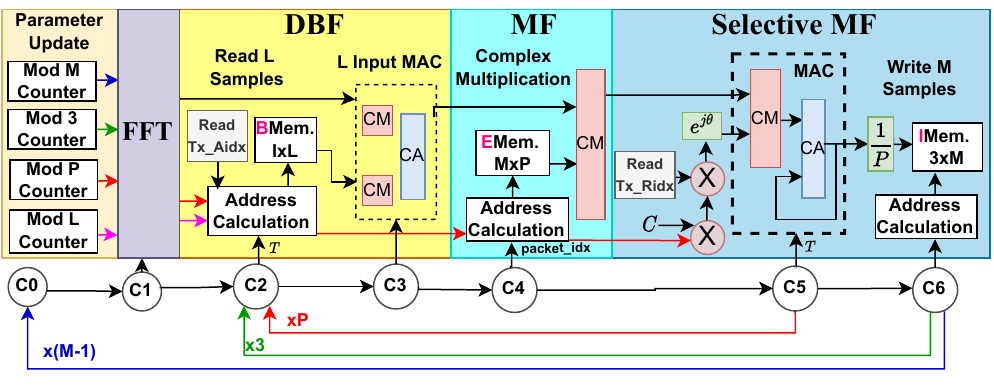}
    \caption{\small Architecture for selective RA processing of multiple packets.}
    \label{fig:sel_mf}
\end{figure}

\vspace{-0.2cm}
\subsection{Doppler Velocity Estimation with MUSIC}
The hardware architectures of SS and MSG are discussed in detail in \cite{tewari2024reconfigrsp}. Figure \ref{fig:givens_rt} presents the hardware architecture of EVD, which uses Givens rotation for QR factorization. As discussed in section \ref{Sec:RSPSysModel}.D, the decomposition of $\mathbf{U}\in\mathbb{C}^{K\times K}$ is achieved through a series of $B=\frac{K(K-1)}{2}$ Givens rotations. Each $\beta=1\cdots B$ rotation involves multiplying $\mathbf{U}^{{\beta-1}}\in\mathbb{C}^{K\times K}$ with a rotation matrix, $\mathbf{Q}^{{\beta}}\in\mathbb{C}^{K\times K}$ to null element $\mathbf{U}^{{\beta-1}}[\mu, \nu]$. Here, $\mu$ and $\nu$ are the row and column indices, respectively, such that $\mu > \nu$. After $B$ iterations, the upper triangular matrix, $R$ is formed as shown,
\vspace{-0.2cm}
\begin{align}
\label{eq:givens}
\small
\mathbf{R}=\mathbf{Q}^{\{1\}}\mathbf{Q}^{\{2\}}\cdots \mathbf{Q}^{\{B\}} \mathbf{U}
\end{align}\normalsize
 The matrix elements of $\mathbf{Q}^{\{\beta\}}$ for the $\beta^{th}$ rotation are, 
 \vspace{-0.2cm}
\begin{align}
\small
\mathbf{Q}^{\{\beta\}}[a,b] =
\begin{cases} 
\cos{\theta}, & [a,b] = [\mu,\mu] \text{ or } [\nu,\nu] \\
-\sin{\theta}, & [a,b] = [\mu,\nu] \\
\sin{\theta}, & [a,b] = [\nu,\mu] \\
1, & a = b \text{ and } [a,b] \notin \{[\mu,\mu], [\nu,\nu]\} \\
0, & a \neq b
\end{cases}
\end{align}\normalsize
Here, $a$ and $b$ denote the indices across the rows and columns of matrix $\mathbf{Q}^{\{\beta\}}$, $\theta$ is the angle between the matrix elements $\mathbf{U}^{{\beta-1}}[\mu, \mu]$ and $\mathbf{U}^{{\beta-1}}[\mu, \nu]$, $\cos{\theta}=\frac{\mathbf{U}^{\{\beta-1\}}[\nu,\nu]}{\sqrt{\mathbf{U}^{\{\beta-1\}}[\nu,\nu]^2+\mathbf{U}^{\{\beta-1\}}[\mu,\nu]^2}}$, and $\sin{\theta}=\frac{\mathbf{U}^{\{\beta-1\}}[\mu,\nu]}{\sqrt{\mathbf{U}^{\{\beta-1\}}[\nu,\nu]^2+\mathbf{U}^{\{\beta-1\}}[\mu,\nu]^2}}$.
 
 In the hardware implementation shown in Figure. \ref{fig:givens_rt}, $\mathbf{U}$ is stored in BRAM $J$, and BRAM $K$ is initialized with the identity matrix. In stage $C1$, based on the value of the counter, the $\mu$ and $\nu$ indices are calculated, and the elements $\mathbf{U}[\nu,\nu]$, and $\mathbf{U}[\mu,\nu]$ are read from BRAM $J$ and stored in registers $u_1$ and $u_2$ respectively. This is followed by the $\cos{\theta}$ and $\sin{\theta}$ calculation in $C2$. The complex square root term 
 in the $\cos{\theta}$ and $\sin{\theta}$ expressions, can be expressed in polar form, as $re^{j\alpha}$. After further refinement, it reduces to ($\sqrt{\frac{r+Re({\mathbf{U}^{\{\beta-1\}}[\nu,\nu]^2+\mathbf{U}^{\{\beta-1\}}[\mu,\nu]^2})}{2}}+j\sqrt{\frac{r-Re({\mathbf{U}^{\{\beta-1\}}[\nu,\nu]^2+\mathbf{U}^{\{\beta-1\}}[\mu,\nu]^2})}{2}}$), facilitating a simplified hardware implementation, as shown in Figure \ref{fig:givens_rt}. B. 
 
 Equation \eqref{eq:givens} involves ($K \times K$) matrix multiplications, $B$ times. These computations can be significantly reduced if each transformation is applied only on the selected $\mu^{th}$ and $\nu^{th}$ rows instead of complete matrix multiplication. In stage $C3$, two samples are read, each from the $\mu^{th}$ and $\nu^{th}$ rows of $\mathbf{U}^{\{\beta-1\}}$ from BRAM $J$. These samples are passed to $C4$ for the Givens Rotation. The resultant transformed samples are then rewritten at the same location in BRAM $J$ in stage $C5$. This is repeated for all the $K$ elements of the selected $\mu^{th}$ and $\nu^{th}$ rows. After this, an identical transformation (stages $C3-C5$) is applied to the $\mu^{th}$ and the $\nu^{th}$ rows of the identity matrix in BRAM $K$. 
 After every $\beta^{th}$ iteration, BRAM $J$ and $K$ are updated with $\mathbf{U}^{\{\beta-1\}}$ and $\mathbf{Q}^{\{\beta\}}$, respectively. In the next iteration, the next set of elements from $\mathbf{U}^{\{\beta-1\}}$ are stored in registers $u_1$ and $u_2$, and the values of $\cos{\theta}$ and $\sin{\theta}$ are updated for the next Givens Rotation. After $B$ iterations, BRAM $J$ is updated with matrix $\mathbf{R}$, and BRAM $K$ is updated with matrix $\mathbf{Q}$, as $\mathbf{Q}=(\mathbf{Q}^{\{1\}}\mathbf{Q}^{\{2\}}\cdots\mathbf{Q}^{\{B\}})^T$. 
 
\begin{figure}[!t]
    \centering
\includegraphics[scale = 0.7]{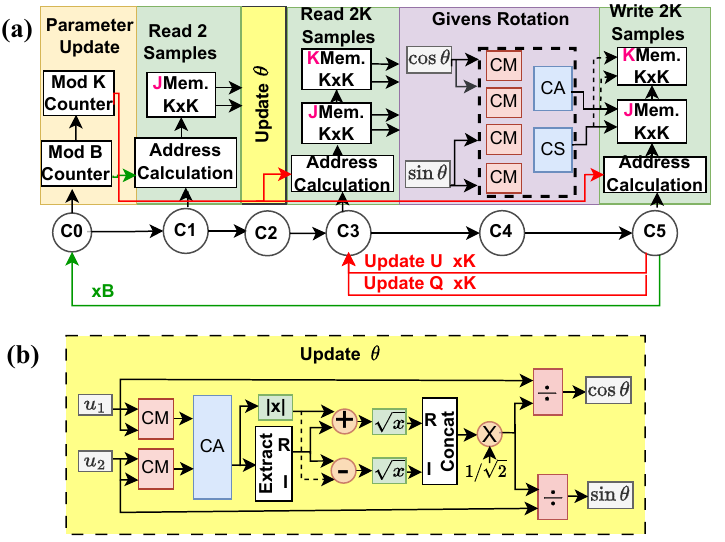}
 \vspace{-0.2cm}
    \caption{\small (a)Hardware architecture for QR factorization using Givens Rotation, (b) $\theta$ calculation block.}
    \label{fig:givens_rt}
     \vspace{-0.2cm}
\end{figure}
\vspace{-0.2cm}
\subsection {Reconfigurable RSP Architecture}
\begin{figure*}[!h]
    \centering
\includegraphics[scale = 0.54]{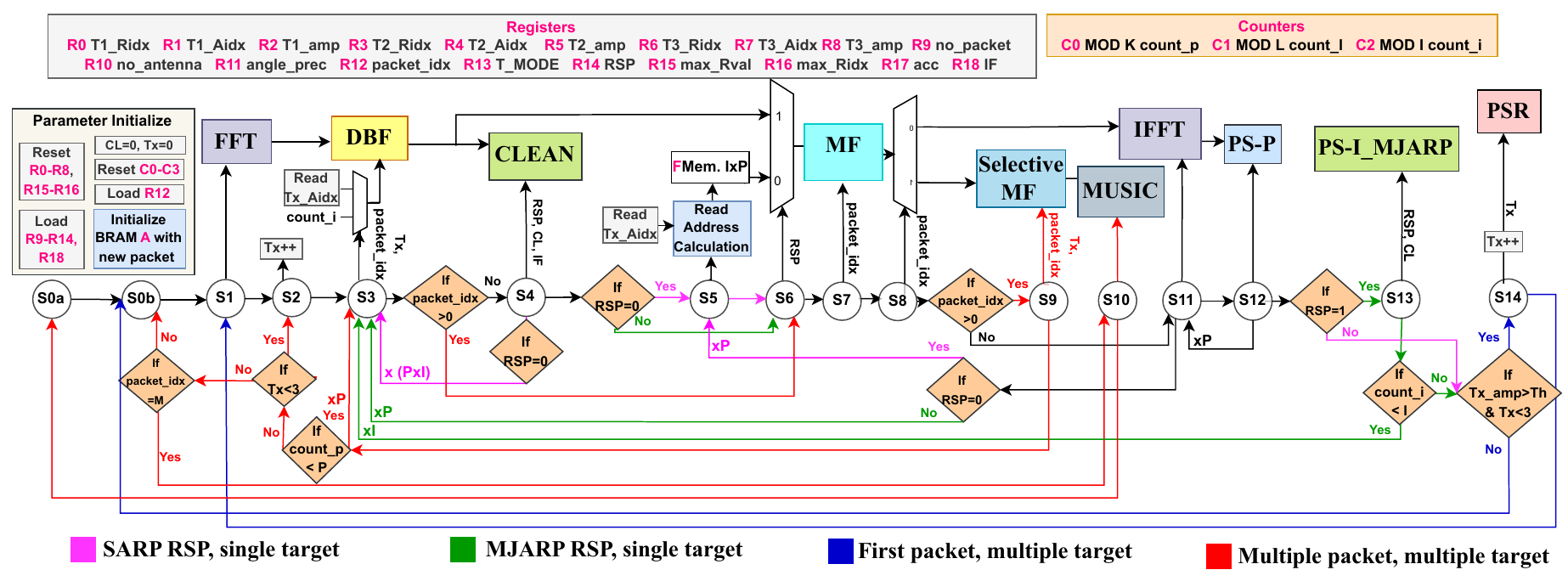}
    \caption{\small Reconfigurable architecture for range, azimuth, and Doppler estimation of multiple targets. (The FFT, DBF, MF, and Selective MF blocks are from Fig.\ref{fig:sel_mf}; IFFT from Fig.~\ref{fig:MF_2D}; CLEAN from Fig.~\ref{fig:clean}; PS-P and PS-I\_MJARP blocks from Fig.~\ref{fig:ps}; PSR from Fig.~\ref{fig:psr}).}
    \label{fig:rsp_hw}
     \vspace{-0.2cm}
\end{figure*}
The reconfigurable architecture that can switch between SARP and MJARP is presented in Figure~\ref{fig:rsp_hw}. 
Since each input may consist of multiple targets, switch between algorithms can happen when new input arrives or when starting the detection of new target in an input. {\color{black} Furthermore, this reconfiguration does not utilize the FPGA's dynamic function exchange (DFX), making it compatible with an ASIC platform. All hardware blocks are vendor-independent and can be synthesized on different FPGAs and ASICs.}

During state $S0$, all registers and counters are first reset, and the RSP configuration parameters are loaded into the PL registers by PS. The first complex packet, $\mathbf{X}_{0}$, is written in BRAM $A$ along with the packet index in $R12$. The counters $C0-C2$ are used for the BRAM read/write address calculation in subsequent stages. Based on the values of $RSP$, $M$, $T$, $\Delta \phi$, $IF$, and $packet\_idx$ registers, the different hardware blocks are configured and reused between different stages. 

First, FFT is performed in $S1$, followed by DBF in $S3$. The DBF unit can be configured to work for a particular search angle (selective RA processing for $m>0$) or for all $I$ search angles. Depending on the packet index, branching occurs at $S3$. Since $packet\_idx=0$ initially, control arrives at $S4$, where the DBF output is written in BRAM $F$ as shown in Figure \ref{fig:clean}. Another branching occurs at state $S4$, based on the $RSP$ flag. Here, state $S5$ is executed only in the case of SARP, where the $P$ selected samples after fast time averaging and azimuth peak search, $\mathbf{Y}_{0,\hat\phi^{\{n\}}}$, are read from BRAM $F$ and passed ahead for MF in $S7$. In the case of MJARP, the control directly arrives at $S6$ from $S4$. Here, all $P\times I$ samples after DBF are sent for MF one by one in a pipelined manner. At state $S8$, branching occurs based on the value of the packet index. For the $packet\_idx=0$, a $P$-point IFFT is performed in $S11$ and a peak search across fast time samples in $S12$. For SARP RSP, registers $R0-R2$ are updated with $T_1$'s peak at the end of this stage. However, for MJARP, an additional peak search across angles is conducted in $S13$. The states $S12$ and $S13$ use the peak search units discussed in Figure \ref{fig:ps}. 

Once the $T_1$ is localized with SARP/MJARP RSP, we localize the next target, $T_2$, with CLEAN. Based on $RSP$, at $S12$/$S13$, if the amplitude of the detected target is below the detection threshold, the control returns to $S0$; otherwise, it arrives at $S14$ for PSR generation. The PSR unit updates BRAM $A$ after incorporating the range and azimuth delays from the previous target, as depicted in Figure \ref{fig:psr}. After this, the control returns to $S1$ for FFT and DBF in $S3$. In stage $S4$, the CLEAN block is configured for complex subtraction of the PSR from the original image. The residue stored in BRAM $F$ is further processed with SARP or MJARP RSP. 

In $S0$, the new packet, $\mathbf{X}_{m}$ is written in BRAM $A$, and the packet index is loaded in $R12$. This is followed by selective RA processing for the slow-time vector creation for each target. After FFT in $S1$, the DBF unit is configured for only the $T_1$'s detected azimuth in $S3$. Since the packet index in $R12$ is greater than zero, the control arrives at state $S9$ for range-index-based IFFT as shown in Figure \ref{fig:sel_mf}. For every packet index, this process is repeated for each target. After $M$ iterations, the BRAM $I$ is updated with the slow-time vectors for all targets. In state $S{10}$, the Doppler estimation is performed sequentially through MUSIC for each target.

\section{Functional Accuracy of Proposed Algorithms}
\label{sec:RSP_PerAnal}
In this section, we analyze the functional accuracy of SARP, MJARP, and reconfigurable RSP approaches by evaluating the root mean square errors (RMSE) in the range and azimuth estimation of multiple targets.
We consider an ISAC BS transceiver with $L=32$ antennas, $P=1024$ fast time samples, PRI of 0.5${\mu}s$ resulting in a maximum detectable radar range, $r_{max}$ of 40 m, and a radar field-of-view spanning -90$^\circ$ to 90$^\circ$ in azimuth. The radar range resolution is 0.085 m due to a radar bandwidth of 1.76 GHz, and the precision for angular search, $\Delta\phi$ is 1$^\circ$. We benchmark the performance of SARP and MJARP RSP over 200 Monte Carlo (MC) experiments. We consider three point targets positioned randomly within the radar field of view with RCS, $\sigma_t$, sampled from Swirling-1 exponential random distribution. The largest point target, $T_1$, has a mean RCS, $<\sigma_t>$, of 10 square meters, which is typically associated with a car. The remaining two point targets, $T_2$ and $T_3$, have mean RCS of 5 and 1 square meters, corresponding to small vehicles (motorbicycles and bicycles) and pedestrians, respectively. We consider a Rician channel with $\ricK=2$ dB to model a strong NLOS component with high multipath, typical of urban environments.
The system SCNR is defined as $\frac{\mathcal{P}_s}{\mathcal{P}_n+\mathcal{P}_c}$ where, $\mathcal{P}_s=\frac{A^2<\sigma_{t}>}{r_{max}^4}$, is the minimum detectable signal strength of the radar. The AWGN noise is modeled as a complex Gaussian distribution with mean noise power, $\mathcal{P}_n$, across all MC iterations. The radar's functional performance is analyzed for different noise floor levels, $\mathcal{P}_n$. The mean clutter power is kept constant at $\mathcal{P}_c = -98dBm$ across all instances. Thus, in each instance of a MC simulation, the SCNR for $T_1$, $T_2$, and $T_3$ is different since each target has a different RCS and is situated at a different distance from the radar but experiences identical noise floors. 

Figure~\ref{fig:rmse} shows the range and azimuth RMSE for the localization of three targets under different noise conditions. 
The MJARP RSP, indicated by the blue solid line, provides the best localization accuracy across all noise levels. The black solid curve indicates the SARP RSP performance under a fast time IF of 512. 
In the case of $T_1$, as shown in Figure \ref{fig:rmse}(a), the SARP RSP shows identical performance as MJARP RSP for noise floors lower than -78 dBm. On the other hand, the MJARP RSP is suitable for higher noise floors.
The red dotted curve highlights the RMSE with reconfigurable architecture that switches from SARP (IF:512) to MJARP at -78 dBm noise floor for $T_1$. For $T_2$, as shown in Figure \ref{fig:rmse}(b), since the scattered signal strength of $T_2$ is weaker as compared to $T_1$ due to weaker RCS, the SARP RSP performance starts deteriorating even at a lower noise floor of -88 dBm compared to $T_1$. Thus, the reconfigurable architecture (shown by the red dotted line) switches from SARP to MJARP RA localization for $T_2$ at -88 dBm. Similarly, in the case of weaker $T_3$, 
the performance deteriorates even at -98 dBm noise floor as highlighted in Figure \ref{fig:rmse}(c). 
The RA localization performance of the reconfigurable architecture is nearly same as the best possible JARP/MJARP across all targets.    
\begin{figure}[!t]
    \centering
    \includegraphics[scale = 0.27]{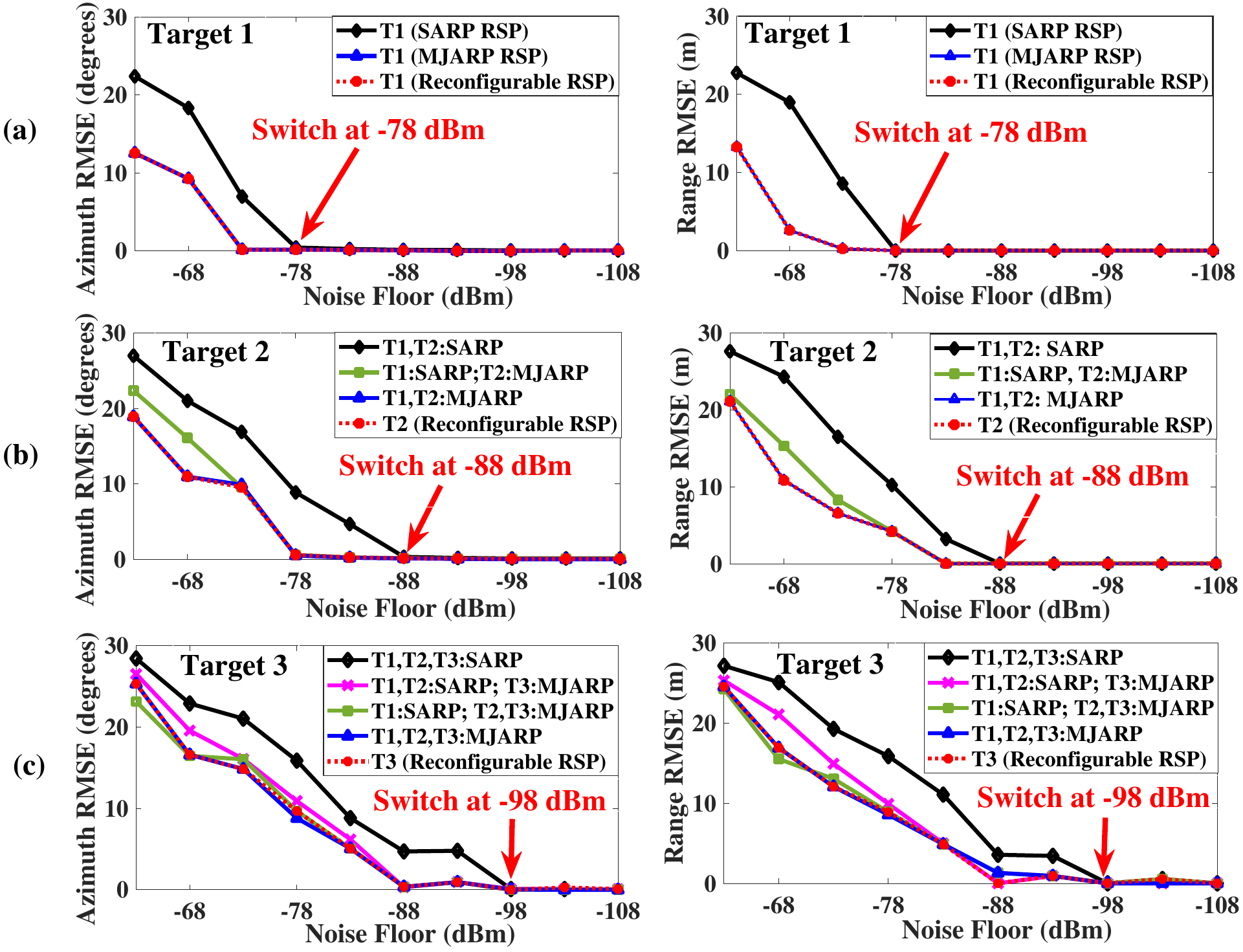}
     \vspace{-0.2cm}
    \caption{\small Azimuth and range RMSE for the detection of 3 targets highlighting switching points, under Rician channel with $\ricK=2$ dB.}
     \vspace{-0.2cm}
    \label{fig:rmse}
\end{figure}
\section{Hardware Complexity Analysis}
\label{Sec:hw_cmpl}
We analyze the complexity of the SARP, MJARP, and reconfigurable RSP on Zynq MPSoC. 
\subsection{Range-Azimuth localization}
\subsubsection{Time-domain vs Frequency-domain matched filtering architecture }
The TD-MF and FD-MF architectures for SARP RSP, discussed in Section.\ref{Sec:1DRSP_arch}, are compared in Table.\ref{tab:1D_TD_FD} in terms of resource utilization and latency. Here, we present two TD-MF configurations- serial and parallel MAC operations, as discussed in Section.\ref{Sec:1DRSP_arch}.1. The execution time reported here is for single target, first packet processing under SARP RSP. The FD-MF implementation requires more resources but is faster than the TD-MF. Hence, FD-MF is used hereafter.  

\begin{table}[!b]
\vspace{-0.2cm}
\caption{\small Hardware complexity comparison of TD-MF and FD-MF for SARP}
\label{tab:1D_TD_FD}
\renewcommand{\arraystretch}{1.05}
\large
\resizebox{\linewidth}{!}{
\centering
\renewcommand{\arraystretch}{1.3}
\begin{tabular}{|c|c|c|c|c|c|c|}
\hline
\textbf{SARP} & \textbf{LUT} & \textbf{FF} & \textbf{BRAM} & \textbf{DSP} & \textbf{Power (W)} & \textbf{Latency (ms)} \\
\hline
\textbf{TD-MF (Parallel)} & 14677 & 20984 & 37 & 44 & 3 & 19.46  \\
\hline
\textbf{TD-MF (Serial)} & 19238 & 44173 & 17 & 16 & 2.98 & 21.45  \\
\hline
\textbf{FD-MF} & 15225 & 37674 & 37 & 88 & 2.95 & 10.75 \\
 & (+3.7\%) & (+80\%) & & (+100\%) & (-1.5\%) & (-48.6\%)  \\
\hline
\end{tabular}}
\end{table}

\subsubsection{Fixed-Point Architecture}
We develop SPFL and fixed-point (FP) hardware IPs for SARP RA localization. 
Here, the functional correctness of the SARP RSP hardware IP (SPFL and FP) is benchmarked against the double precision floating point (DPFL) implementation in PS. In Figure~\ref{fig:wl_dbf} and Figure~\ref{fig:wl_mf}, we evaluate the range-azimuth RMSE under different noise floor levels for various word lengths (WL). Figure~\ref{fig:wl_dbf} shows the performance of DBF at different WLs with MF in DPFL. The SPFL implementation of SARP, indicated by a black dashed line, offers the same RMSE as its DPFL implementation, shown with a red solid line for all noise floors. Therefore, there is no degradation in functional accuracy while transitioning from DPFL to SPFL. Next, we consider the FP representation as $<W,Z>$, where $W$, $Z$, and $(W-Z)$ represent the total number of bits, number of integer bits, and number of fractional bits, respectively. First, we fix $Z=2$, by appropriately scaling the input such that 2 integer bits are sufficient for the signed representation, and compare RMSE for different values of $W$. Figure~\ref{fig:wl_dbf} shows that $W=22$ has the same RMSE as that of DPFL. Further, the azimuth RMSE deviates slightly for $W=21$, and the error propagates to subsequent range estimation. Thus, WL of $<22,2>$ is selected for DBF.
\begin{figure}[!ht]
    \centering
    \includegraphics[scale = 0.4]{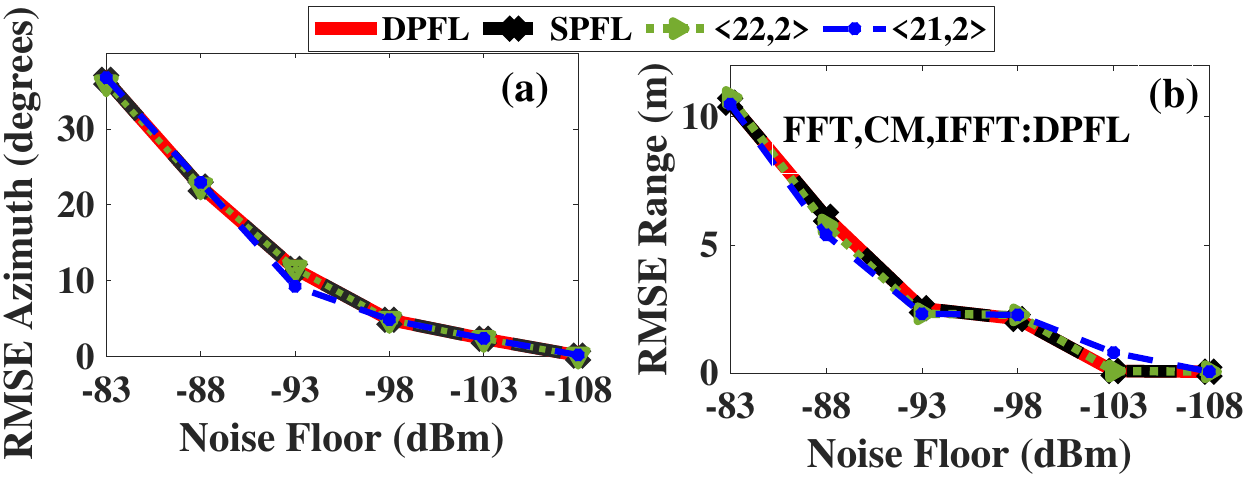}
     \vspace{-0.2cm}
    \caption{ \small{WL selection for DBF with  MF (FFT, CM, IFFT) in DPFL.}}
     \vspace{-0.1cm}
    \label{fig:wl_dbf}
\end{figure}
\begin{figure*}[!ht]
    \centering
    \includegraphics[scale = 0.4]{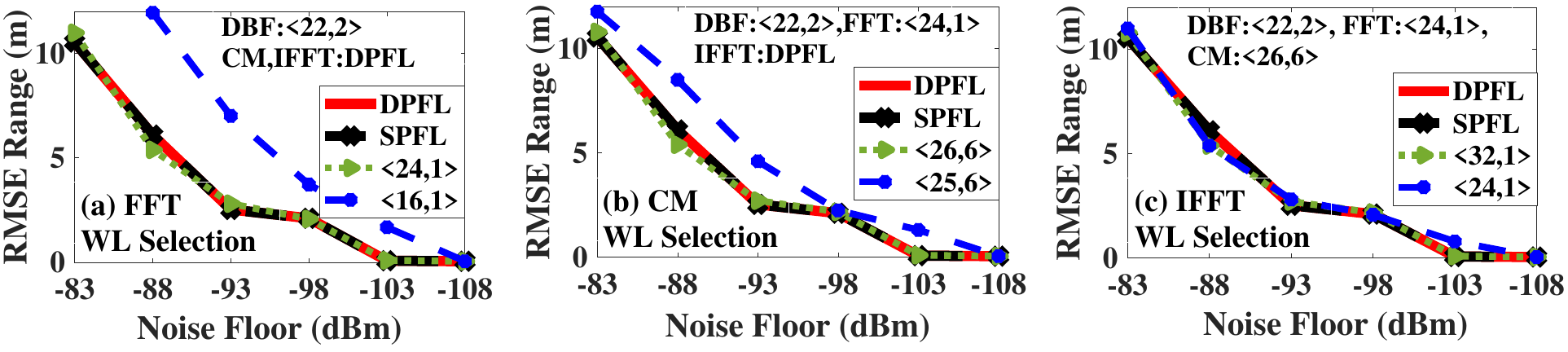}
    \vspace{-0.2cm}
    \caption{ \small Word-length selection for MF blocks, (a) FFT, (b) CM, (c) IFFT with DBF in $<$22,2$>$ FP architecture.}
    \label{fig:wl_mf}
     \vspace{-0.2cm}
\end{figure*}
Next, we discuss the WL for MF, which comprises FFT, CM, and IFFT. Figure~\ref{fig:wl_mf}(a) shows the WL selection of the FFT block, with CM and IFFT in DPFL. The inbuilt AMD Xilinx IP for FFT/IFFT offers only selected FP architectures with WL: 16, 24, and 32. As shown in Figure~\ref{fig:wl_mf}(a), the WL $<24,1>$ is the appropriate choice for FFT, showing RMSE equivalent to DPFL. 
With FFT WL of $<24,1>$, and IFFT in DPFL, the WL for CM is selected as $<26,6>$, and the corresponding RMSE comparison is shown in Figure~\ref{fig:wl_mf}(b). In the same manner, WL of $<32,1>$ is selected for IFFT, as shown in Figure~\ref{fig:wl_mf}(c). 

Table~\ref{tab:wl_hw_cmpl} compares the resource utilization, power consumption, and execution time for SPFL and FP SARP architecture on Zynq MPSoC. These results are for single target and single packet SARP and do not include the hardware complexity of CLEAN. The FP implementation with DBF:$<22,2>$, FFT:$<24,1>$, CM:$<26,6>$, IFFT:$<32,1>$ provides a significant reduction, up to 40-70\%, in resource utilization and 40\% in power consumption for both 8 and 32 antenna architectures. Also, the FP SARP RSP is 45\% faster than SPFL. Therefore, the FP design results in lower hardware cost and faster execution time compared to SPFL, without any degradation in RA estimation performance, as highlighted in Figures.~\ref{fig:wl_dbf} (a) and \ref{fig:wl_mf} (c), respectively.
\begin{table}[h]
\caption{\small SARP single target hardware complexity comparison with SPFL and FP WL for different antenna numbers.}
\label{tab:wl_hw_cmpl}
\huge
\renewcommand{\arraystretch}{1.05}
\resizebox{\linewidth}{!}{
\centering
\begin{tabular}{|c|c|c|c|c|c|c|c|}
\hline
\textbf{\begin{tabular}[c]{@{}c@{}}No. of \\ Antennas\end{tabular}} & \textbf{Architecture} & \textbf{LUT} & \textbf{FF} & \textbf{BRAM} & \textbf{DSP} & \textbf{Power (W)} & \textbf{Latency (ms)} \\
\hline
\multirow{2}{*}{\textbf{8}}  & SPFL & 26203 & 49999 & 65 & 242 & 3.55 & 14.74 \\ \cline{2-8}
& FP & 16420 & 35340 & 32.5 & 86 & 2 & 8 \\
& & (-37.3\%) & (-29.4\%) & (-50\%) & (-64\%) & (-44\%) & (-45.75\%) \\
\hline
\multirow{2}{*}{\textbf{32}} & SPFL & 58033 & 73923 & 149 & 722 & 3.17 & 15.42 \\ \cline{2-8}
& FP & 20278 & 20072 & 119.5 & 294 & 2.02 & 8.26 \\
& & (-65\%) & (-72.8\%) & (-19.8\%) & (-59.2\%) & (-36.8\%) & (-46.43\%) \\
\hline
\end{tabular}}
\end{table}
The FP selection for MJARP is performed in an identical manner to SARP with the selected WL as DBF:$<22,2>$, FFT:$<24,1>$, CM:$<26,6>$, IFFT:$<32,1>$. 

\subsubsection{Comparison between JARP, MJARP, SARP, and reconfigurable RSP} Table~\ref{tab:rsp_hwcmp} compares the hardware complexity for  RA localization of three targets with JARP, MJARP, SARP, and reconfigurable architectures. These architectures are implemented with FP MF and DBF, along with CLEAN in SPFL. Here, the FP complexity of SARP is obtained from the previous discussion. To compare the execution time, we use the metric acceleration factor (AF), which is the ratio of the execution time of the given architecture to the baseline execution time of JARP reported in \cite{tewari2024reconfigrsp}. Due to the longest execution time, the AF of JARP is taken as 1. Here, SARP and MJARP both show higher AF and additional resource utilization than JARP due to the shifting of peak search and CLEAN to PL. As expected, MJARP is computationally more complex in terms of resources and power requirements than SARP and requires higher execution time due to multiple calls to FFT and IFFT cores for MF. 
\textit{The reconfigurable SARP-MJARP RSP architecture offers a maximum AF of 8 without any additional hardware complexity compared to MJARP RSP.} The AF improves significantly for lower IF. 

\begin{table}[!b]
\caption{\small Hardware complexity comparison of RSP architectures for three target RA localization on Zynq MPSoC. }
\label{tab:rsp_hwcmp}
\renewcommand{\arraystretch}{1.05}
\large
\resizebox{\linewidth}{!}{
\centering
\begin{tabular}{|c|c|c|cccc|c|}
\hline
\multirow{2}{*}{\textbf{RSP}} &
  \multirow{2}{*}{\textbf{\begin{tabular}[c]{@{}c@{}}Fast time\\ I.F.\end{tabular}}} &
  \multirow{2}{*}{\textbf{A.F.}} &
  \multicolumn{4}{c|}{\textbf{FPGA Resource Utilization}} &
  \multirow{2}{*}{\textbf{\begin{tabular}[c]{@{}c@{}}Power\\  (W)\end{tabular}}} \\ \cline{4-7}
 &
   &
   &
  \multicolumn{1}{c|}{\textbf{LUT}} &
  \multicolumn{1}{c|}{\textbf{FF}} &
  \multicolumn{1}{c|}{\textbf{BRAM}} &
  \textbf{DSP} &
   \\ \hline
   JARP \cite{tewari2024reconfigrsp} &
  1 &
  1 & \multicolumn{1}{c|}{21336} & \multicolumn{1}{c|}{24703} & \multicolumn{1}{c|}{381} & \multicolumn{1}{c|}{300} & \multicolumn{1}{c|}{2.8}\\ \hline 
  \multirow{2}{*}{MJARP} & 1 & 2.15 & \multicolumn{1}{c|}{\begin{tabular}[c]{@{}c@{}}33231\\ (+55.7\%)\end{tabular}} & \multicolumn{1}{c|}{\begin{tabular}[c]{@{}c@{}}29295\\(+18.5\%)\end{tabular}} & \multicolumn{1}{c|}{\begin{tabular}[c]{@{}c@{}}451\\(+18.3\%)\end{tabular}} & \multicolumn{1}{c|}{\begin{tabular}[c]{@{}c@{}}412\\(+37.3\%)\end{tabular}} & \begin{tabular}[c]{@{}c@{}}2.9\\(+2.8\%)\end{tabular}\\ \hline
\multirow{3}{*}{SARP} &
  1024 &
  3.01 &
  \multicolumn{1}{c|}{\multirow{3}{*}{\begin{tabular}[c]{@{}c@{}}28315 \\(+32.7\%)\end{tabular}}} &
  \multicolumn{1}{c|}{\multirow{3}{*}{\begin{tabular}[c]{@{}c@{}}24664 \\(-0.15\%)\end{tabular}}} &
  \multicolumn{1}{c|}{\multirow{3}{*}{\begin{tabular}[c]{@{}c@{}}189.5 \\(-50.3\%)\end{tabular}}} &
  \multirow{3}{*}{\begin{tabular}[c]{@{}c@{}}406 \\(+35.3\%)\end{tabular}} &
  \multirow{3}{*}{\begin{tabular}[c]{@{}c@{}}2.34 \\(-16.4\%)\end{tabular}} \\ \cline{2-3}
 &
512 &
  4.3 &
  \multicolumn{1}{c|}{} &
  \multicolumn{1}{c|}{} &
  \multicolumn{1}{c|}{} &
   &
   \\ \cline{2-3}
 &
  64 &
  8 &
  \multicolumn{1}{c|}{} &
  \multicolumn{1}{c|}{} &
  \multicolumn{1}{c|}{} &
   &
   \\ \hline
\textbf{\begin{tabular}[c]{@{}c@{}}Reconfigurable\\ Architecture\end{tabular}} &
  \textbf{512/1024} &
  \begin{tabular}[c]{@{}c@{}}\textbf{upto} \\\textbf{8}\end{tabular} &
  \multicolumn{1}{c|}{\begin{tabular}[c]{@{}c@{}}33231\\ \textbf{(+55.7\%)}\end{tabular}} &
  \multicolumn{1}{c|}{\begin{tabular}[c]{@{}c@{}}29295\\ \textbf{(+18.5\%)}\end{tabular}} &
  \multicolumn{1}{c|}{\begin{tabular}[c]{@{}c@{}}451\\ \textbf{(+18.3\%)}\end{tabular}} &
  \begin{tabular}[c]{@{}c@{}}412 \\ \textbf{(+37.3\%)}\end{tabular} & \textbf{2.3-2.9}
   \\ \hline
\end{tabular}}
\end{table}

\subsection{Doppler velocity Estimation}
We perform the Doppler estimation using MUSIC algorithm with different architectures on Zynq MPSoC and benchmark the hardware complexity against Doppler estimation using FFT in Table.~\ref{tab:doppler_hw}. For an ideal comparison between FFT and MUSIC, we keep the same number of slow-time input samples (32) and Doppler precision (0.3 m/s) across both algorithms. The FFT core is developed with FP WL $<24,1>$ using the inbuilt AMD Xilinx LogiCORE FFT IP and shows the Doppler estimation accuracy very close to SPFL. The FFT FP has the least hardware complexity and execution time among all implementations, making it a preferred choice. However, FFT does not perform well when two or more targets have to be resolved within the same range-azimuth bin. MUSIC offers super-resolution capability and is preferred for such cases. Figure \ref{fig:music_rmse} compares the Doppler velocity RMSE under super-resolution conditions across different MUSIC hardware architectures. The red dotted curve shows MUSIC using the in-built AMD Xilinx QRF IP for EVD. Here, EVD is in SPFL and fails to resolve two targets in the same RA cell with Doppler velocities 4 m/s apart. {\color{black} The vendor QRF IP in \cite{tewari2024reconfigrsp} supports only SPFL and is functionally unsuitable since it is prone to not-a-number generation under high dynamic range conditions.} Our custom hardware IP for EVD, based on Givens Rotation, has all the blocks- SS, EVD, and MSG in DPFL, resulting in high resource utilization and latency, as shown in Table~\ref {tab:doppler_hw}. As shown by the blue solid line in Figure \ref{fig:music_rmse}, this approach is able to resolve the two targets. \textit{Implementing only the MSG unit with FP $<32,12>$ architecture and keeping the SS and EVD in DPFL reduces the latency of Doppler estimation by 85\%, along with a reduction in resources compared to the DPFL MUSIC implementation. Additionally, this implementation also provides a similar performance to the DPFL}, as shown by the black solid line in Figure \ref{fig:music_rmse}. Thus, this FP implementation of MUSIC is most suitable for super-resolution scenarios. 
\begin{figure}[!t]
    \centering
    \includegraphics[scale = 0.4]{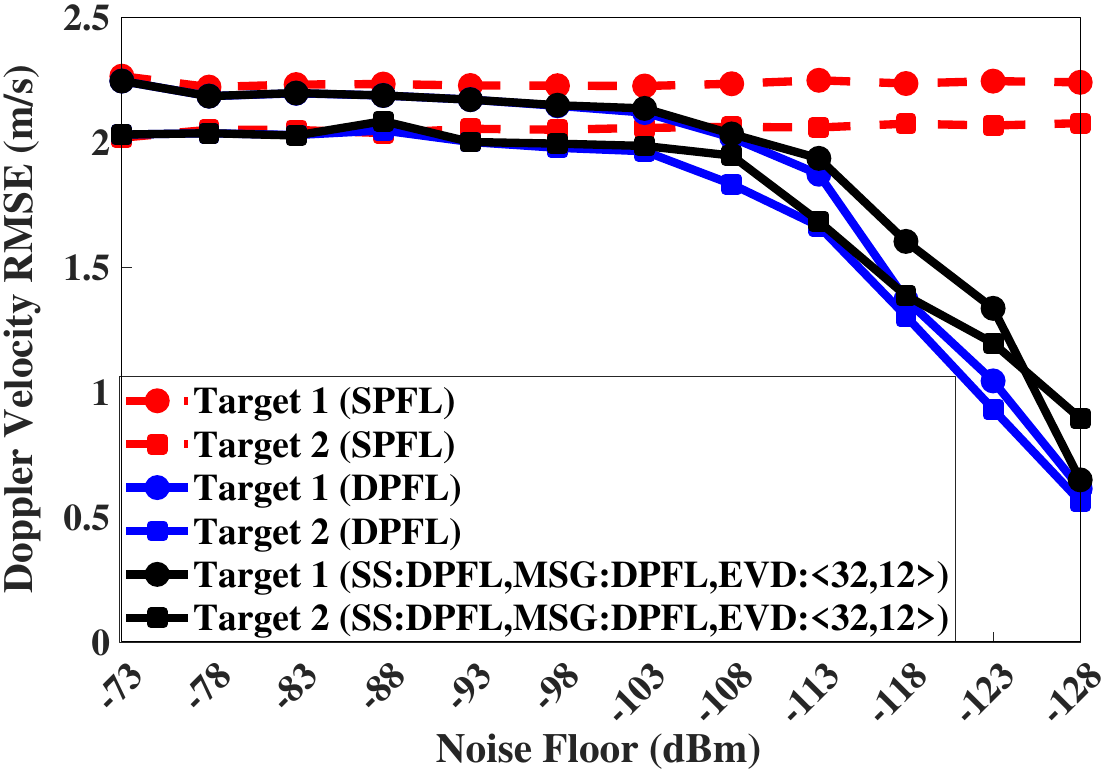}
    \caption{\small Doppler velocity RMSE on FPGA, under different WL architectures of MUSIC, showing super-resolution for two targets with a velocity difference of 4m/s, with 32 packets.}
    \label{fig:music_rmse}
\end{figure}

\begin{table}[!t]
\caption{ \small Hardware complexity for Doppler velocity estimation with FFT and MUSIC for 32 packets }
\label{tab:doppler_hw}
\Huge
\renewcommand{\arraystretch}{1.05}
\resizebox{\linewidth}{!}{
\centering
\begin{tabular}{|c|c|c|c|c|c|c|}
\hline
\textbf{\begin{tabular}[c]{@{}c@{}}Doppler \\ Estimation\end{tabular}} &
  \textbf{\begin{tabular}[c]{@{}c@{}}Hardware \\  IP Core\end{tabular}} &
  \textbf{Word-length} &
  \textbf{\begin{tabular}[c]{@{}c@{}}Resource Utilization    \\ \{BRAM, LUT, FF, DSP\}\end{tabular}} &
  \textbf{\begin{tabular}[c]{@{}c@{}}Power \\ (W)\end{tabular}} &
  \textbf{\begin{tabular}[c]{@{}c@{}}Latency \\ (ms)\end{tabular}} &
  \textbf{\begin{tabular}[c]{@{}c@{}}Super- \\ resolution\end{tabular}} \\ \hline
\textbf{FFT} &
  \begin{tabular}[c]{@{}c@{}}AMD Xilinx\\ LogiCORE\end{tabular} &
  \textless{}24,1\textgreater{} &
  72.5,   10872, 15778, 18 &
  3.65 &
  0.52 &
  \xmark \\ \hline
\multirow{5}{*}{\textbf{MUSIC}} &
  \begin{tabular}[c]{@{}c@{}}AMD Xilinx\\ QRF (EVD)\end{tabular} &
  SPFL &
  90, 77643, 83532, 458 &
  3.607 &
  0.99 &
  \xmark \\ \cline{2-7} 
 &
  \multirow{4}{*}{\begin{tabular}[c]{@{}c@{}}Givens \\ Rotation   \\ (EVD)\end{tabular}} &
  DPFL &
  252, 79721, 61717, 723 &
  4.21 &
  5.13 &
  \checkmark \\ \cline{3-7} 
 &
   &
  \begin{tabular}[c]{@{}c@{}}SS:DPFL,\\FP:\textless{}32,12\textgreater{},\\MSG:DPFL\end{tabular} &
  \begin{tabular}[c]{@{}c@{}}90, 78467, 72705, 599   \\ \textbf{(-64.2\%, -1.5\%, +17.8\%, -17\%)}\end{tabular} &
  \begin{tabular}[c]{@{}c@{}}3.696 \\ \textbf{(-12.2\%)}\end{tabular} &
  \begin{tabular}[c]{@{}c@{}}0.75   \\ \textbf{(-85.4\%)}\end{tabular} &
  \checkmark \\ \hline
\end{tabular}
}
\end{table}
\color{black}
\begin{table*}[!t]
 \centering
  \caption{\small HSCD of proposed multiple target, 3D reconfigurable RSP on Zynq MPSoC } 
  \Huge
  \renewcommand{\arraystretch}{1.3}
  \label{tab:hw_cmpl}
 \resizebox{\textwidth}{!}{
\begin{tabular}{|cc|c|c|c|c|c|c
|c|c|c|}
\hline
\textbf{Packets} & \multicolumn{2}{|c|}{\textbf{RSP}}  &
  \multicolumn{1}{c|}{\textbf{PS}} &
  \multicolumn{1}{c|}{\textbf{PL}} &
  \multicolumn{1}{c|}{\textbf{\begin{tabular}[c]{@{}c@{}}Resource Utilization   \\ \{BRAM, DSP, FF, LUT\}\end{tabular}}} &
    \multicolumn{1}{c|}{\textbf{\begin{tabular}[c]{@{}c@{}}Dynamic\\Power (W)\end{tabular}}} &
  \multicolumn{1}{c|}{\textbf{\begin{tabular}[c]{@{}c@{}}Execution \\ Time (ms)\end{tabular}}} &
  \multicolumn{1}{c|}{\textbf{\begin{tabular}[c]{@{}c@{}}AF w.r.t.\\  PS\end{tabular}}} &
  \multicolumn{1}{c|}{\textbf{\begin{tabular}[c]{@{}c@{}}AF w.r.t. \\ JARP \cite{tewari2024reconfigrsp}\end{tabular}}} &
  \textbf{Opt.}\\ \hline
\multirow{6}{*}{\textbf{\begin{tabular}[c]{@{}c@{}}1 \\ (first \\ packet)\end{tabular}}} & \multicolumn{1}{|c|}{\multirow{9}{*}{\textbf{\begin{tabular}[c]{@{}c@{}}Range and \\ Azimuth\end{tabular}}} } & - &
   \begin{tabular}[c]{@{}l@{}} DBF, MF, Peak \\search, CLEAN\end{tabular}  &
  - &
  - &
  - &
  1481 &
  1 &
  - &
  \multirow{5}{*}{a} \\ \cline{3-3} \cline{4-10} & 
\multicolumn{1}{|c|}{} &
 JARP&
  Peak search, CLEAN &
  DBF (FP), MF (FP) &
  381, 300, 24703, 21336 &
  2.8&
  124 &
  11.9 &
  1 &
  \\  \cline{3-3} \cline{4-10} 
\multicolumn{1}{|c|}{} &
   & -
   &
  CLEAN &
  DBF (FP), MF (FP), Peak search (SPFL) &
  382, 367, 23057, 26075 &
  - &
  70 &
  21 &
  1.77 & \\ \cline{3-3}\cline{4-10} 
\multicolumn{1}{|c|}{} & &
 MJARP  &
  - &
  \begin{tabular}[c]{@{}l@{}}DBF (FP), MF (FP), Peak search (SPFL),\\ CLEAN   (SPFL)\end{tabular} &
  451, 412, 29295, 33231 &
  2.9 &
  58 &
  25.5 &
  2.15 & \\ \cline{3-3} \cline{4-11} 
\multicolumn{1}{|c|}{} & &
   \multirow{1}{*}{Reconfigurable}
   &
  - & 
  \begin{tabular}[c]{@{}l@{}}DBF (FP), MF (FP), Peak search (SPFL),\\ CLEAN   (SPFL)\end{tabular} &
  451, 412, 29295, 33231 & 
  2.34-2.9  &
  29-58 &
  25.5-51 &
  2.15-4.3 & b \\ \cline{1-1} \cline{3-11} 
  \multirow{3}{*}{\textbf{32}} & \multicolumn{1}{|c}{}&\multicolumn{1}{|c|}{JARP}& 
  Peak search, CLEAN &
  DBF (FP), MF (FP) &
  381, 300, 24703, 21336 &
  2.8 &
  782 &
  - &
  1 & \multirow{3}{*}{c} \\ \cline{3-3} \cline{4-10} 
\multicolumn{1}{|c|}{} & & Reconfigurable
   &
  - &
  \begin{tabular}[c]{@{}l@{}}DBF (FP), MF (FP), Peak search (FP), \\ CLEAN (SPFL)\end{tabular} &
  \begin{tabular}[c]{@{}c@{}}451, 412, 29295, 33231   \\ \textbf{(+18\%, +37\%, +18.6\%, +55.7\%)}\end{tabular} &
  \begin{tabular}[c]{@{}c@{}}2.34-2.9\\ \textbf{(upto -16.4\%)}\end{tabular} &
  134-163 &
  - &
  4.8-5.62 & \\ \hline
\multirow{5}{*}{\textbf{32}} &
  \multicolumn{1}{|c|}{\multirow{5}{*}{\textbf{\begin{tabular}[c]{@{}c@{}}Range, \\ Azimuth, and\\  Doppler velocity \\ (super-resolution)\end{tabular}}}} &\multirow{1}{*}{JARP} &
  Peak search, CLEAN &
  DBF (FP), MF (FP), MUSIC (DPFL) &
  633, 1023, 86420, 101057 &
  3.3 &
  800 &
  - &
  1 & \multirow{5}{*}{d}\\ \cline{3-3}\cline{4-10} 
& \multicolumn{1}{|c|}{} &
\multirow{3}{*}{Reconfigurable}
   &
  - &
  \begin{tabular}[c]{@{}l@{}}DBF (FP), MF (FP), Peak search (FP), \\ CLEAN (SPFL), MUSIC (DPFL)\end{tabular} &
  \begin{tabular}[c]{@{}l@{}}703, 1135, 91012, 112952 \\ \textbf{(+11\%, +11\%, +5.3\%, +11.8\%)}\end{tabular} &   \begin{tabular}[c]{@{}c@{}}2.76-3.9\\ \textbf{(upto -16.4\%)}\end{tabular} &
  149-178 & 
  - &
  4.49-5.17 & \\ \cline{4-10} 
\multicolumn{1}{|c|}{} &
   &
   &
  - &
  \begin{tabular}[c]{@{}l@{}}DBF (FP), MF (FP), Peak search (FP), \\ CLEAN (SPFL), MUSIC (FP)\end{tabular} &
  \begin{tabular}[c]{@{}l@{}}541, 1011, 102000, 111698\\ \textbf{(-14.5\%,   -1.2\%, +18\%,  +10.5\%) }\end{tabular} & \begin{tabular}[c]{@{}c@{}}2.43-3.43\\ \textbf{(upto -26.4\%)}\end{tabular}
 &
  137-166 &
  - &
  4.82-5.62 &\\ \hline
\end{tabular}}
   \label{tab:hw_azi}
\end{table*}

\begin{table*}[!h]
\caption{\textcolor{black}{Comparison of proposed work with prior works }}
\label{tab:prior_art}
\huge
\renewcommand{\arraystretch}{1}
\resizebox{\linewidth}{!}{
\centering
\color{black}{
\begin{tabular}{|c|c|c|c|c|c|c|c|c|c|c|c|}
\hline
\multirow{2}{*}{\textbf{Reference}} &
  \multirow{2}{*}{\textbf{Platform}} &
  \multirow{2}{*}{\textbf{RSP}} &
  \multirow{2}{*}{\textbf{\begin{tabular}[c]{@{}c@{}}Processing   \\ Dimension\end{tabular}}} &
  \multirow{2}{*}{\textbf{Resolution}} &
  \multirow{2}{*}{\textbf{Fixed-point}} &
  \multicolumn{4}{c|}{\textbf{Resource Utilization}} &
  \multirow{2}{*}{\textbf{\begin{tabular}[c]{@{}c@{}}Latency/   \\ Packet   (ms)\end{tabular}}} &
  \multirow{2}{*}{\textbf{\begin{tabular}[c]{@{}c@{}}Power   \\  (W)\end{tabular}}} \\ \cline{7-10}
 &
   &
   &
   &
   &
   &
  \multicolumn{1}{c|}{\textbf{FF}} &
  \multicolumn{1}{c|}{\textbf{LUT}} &
  \multicolumn{1}{c|}{\textbf{BRAM}} &
  \textbf{DSP} &
   &
   \\ \hline
\cite{schweizer2021fairy}&
  ZCU111 &
  \begin{tabular}[c]{@{}c@{}}range,   \\ Doppler\end{tabular} &
  2D-512 FFT &
  \begin{tabular}[c]{@{}c@{}}0.155$m$,   \\ 0.37$m/s$\end{tabular} &
  SPFL &
  \multicolumn{1}{c|}{NA} &
  \multicolumn{1}{c|}{221402} &
  \multicolumn{1}{c|}{NA} &
  1587 &
  102.4 &
  NA \\ \hline
\cite{Lin2024HWspaceborneSAR} &
  \begin{tabular}[c]{@{}c@{}}Virtex    \\ VCU128\end{tabular} &
  \begin{tabular}[c]{@{}c@{}}range,    \\ azimuth\end{tabular} &
  2D-8192 FFT &
   \begin{tabular}[c]{@{}c@{}}10$m$,   \\ 0.2$^\circ$\end{tabular} &
  FFT: 17 bit &
  \multicolumn{1}{c|}{208415} &
  \multicolumn{1}{c|}{537034} &
  \multicolumn{1}{c|}{1423} &
  252 &
  340 &
  27.2 \\ \hline
\cite{zhang2023HW_EVD_tvt} &
  Zedboard &
  azimuth &
  3 antennas &
  57$^\circ$ &
  \begin{tabular}[c]{@{}c@{}}COV:\textless{}30,2\textgreater   \\ EVD:\textless{}15,12\textgreater{}\end{tabular} &
  \multicolumn{1}{c|}{7983} &
  \multicolumn{1}{c|}{7086} &
  \multicolumn{1}{c|}{0.5} &
  36 &
  0.002 &
  0.392 \\ \hline
\cite{li2022musictcas1} &
  Zynq-7000 &
  azimuth &
  8 antennas &
  10$^\circ$ &
  \begin{tabular}[c]{@{}c@{}}COV,EVD:10 bits   \\ MSG:20 bits\end{tabular} &
  \multicolumn{1}{c|}{14228} &
  \multicolumn{1}{c|}{12820} &
  \multicolumn{1}{c|}{596} &
  79 &
  0.012 &
  NA \\ \hline
\multirow{5}{*}{Proposed work} &
  \multirow{5}{*}{ZCU111} &
  \multirow{2}{*} {range}&
  \multirow{2}{*} {1024 FFT} &
  \multirow{2}{*} {0.08$m$} &
  \multirow{4}{*}{\begin{tabular}[c]{@{}c@{}}DBF:\textless{}22,2\textgreater{},   \\  FFT:\textless{}24,1\textgreater{},   \\  CM:\textless{}26,6\textgreater{},   \\  IFFT:\textless{}32,1\textgreater{}\end{tabular}} &\multirow{2}{*} {32416} &\multirow{2}{*} {14271} &\multirow{2}{*} {356} &
  \multirow{2}{*} {156} &
  \multirow{2}{*} {27.3} &
  \multirow{2}{*} {1.73} \\ & & & & & & & & & & &  \\\cline{3-5} \cline{7-12} 
 &
   &
  \multirow{2}{*} {azimuth} &
  \multirow{2}{*} {32 antennas} &
  \multirow{2}{*} {6$^\circ$} &
   &
  \multirow{2}{*} {7399} &
  \multirow{2}{*} {5511} & \multirow{2}{*} {95} &
  \multirow{2}{*} {256} &
  \multirow{2}{*} {1.7} &
  \multirow{2}{*} {0.67} \\ 
  & & & & & & & & & & &\\\cline{3-12} 
 &
   &
  \multirow{1}{*} {Doppler} &
  32 snapshots &
  0.3$m/s$ &
  MSG:\textless{}32,1\textgreater{} &
  \multicolumn{1}{c|}{72705} &
  \multicolumn{1}{c|}{78467} &
  \multicolumn{1}{c|}{90} &
  599 &
  0.75 &
  3.7 \\ \hline
\end{tabular}
}
}
\end{table*}

\subsection{Hardware Software Co-design on Zynq MPSoC} 
We partition the various RSP tasks- RA localization via MF and DBF, peak search, CLEAN for multiple target detection, and MUSIC between PL and PS of Zynq MPSoC via HSCD. This profiling analysis is provided in Table.~\ref{tab:hw_cmpl}. First, we analyze the latency requirement for the RA localization of the first packet, $\mathbf{X}_0$, indicated in the first five rows of the table. The first row shows the entire RSP, including DBF, MF, peak search, and CLEAN, performed in PS for the detection of three targets. This configuration results in the longest execution time as the FPGA is not involved, and hence its AF w.r.t PS is indicated as 1. Next, we shift the DBF and MF blocks to PL, keeping peak search and CLEAN in PS only. This is the JARP configuration, which results in a speed-up of 12 times compared to PS. Next, shifting the peak search unit also to the PL further improves the AF with a slight increase in FPGA resources. This improvement is because the peak search is conducted efficiently in PL, pipelined with the BRAM write operations, as discussed in Section \ref{Sec:RSP_hw}.C. Further, shifting the CLEAN to PL results in an AF of 25 times w.r.t PS and a 2 times improvement in RA localization latency (three targets) over JARP. This is the MJARP configuration, where the entire RSP is offloaded to PL. 
Integration of CLEAN into the PL is the first optimization of the reconfigurable architecture and is referred to as $a$ in Table \ref {tab:hw_cmpl}. The next optimization $b$ of the reconfigurable architecture is the incorporation of SARP RA localization, which further accelerates 4.3 times w.r.t JARP.
\\\indent Next, we analyze the latency for RA processing of 32 packets, including the multiple target processing with CLEAN on the first packet and selective RA processing of the remaining packets. The first configuration is JARP, with DBF and MF in PL, and peak search and CLEAN performed in PS. Here, RA processing of all packets is conducted for $I$ search angles. Our proposed reconfigurable architecture with entire RA localization in PL provides an AF of 5.6 times compared to JARP with $(18-55)\%$ increase in resource utilization. This significant improvement in AF is because the PL to PS transfer of a large number of packets, each of dimension $\Gamma_m \in \mathbb{C}^{I \times P}$, is avoided. Additionally, only the selective RA processing is performed for all packets except the first packet. This optimization in the reconfigurable architecture for the RA localization of multiple packets is referred to as $c$ in the table. Next, we benchmark the complete 3D RSP with 32 packets for the proposed reconfigurable architecture termed as $d$ (DBF, MF, peak search, CLEAN, and MUSIC in PL) against JARP, where only DBF, MF, and MUSIC are in PL. Here, we design a Doppler velocity estimation with super-resolution capability. As discussed previously, MUSIC must be performed in DPFL for super-resolution. Thus, we first compare the reconfigurable RSP and JARP for DPFL MUSIC. As expected, the reconfigurable architecture provides an AF of up to 5 times over JARP, with only 11\% increase in FPGA resources. The proposed FP MUSIC architecture (SS and EVD in DPFL and MSG in $<32,12>$) offers a reduction in resources and an improvement in latency over DPFL MUSIC. \textit{Thus, the reconfigurable architecture with 3D processing (range, azimuth, and Doppler) provides an AF of up to 5.6 times over JARP for 32 packets. This is achieved with fewer DSP and memory units and up to 26.4\% dynamic power consumption reduction.} 

\textcolor{black}{ We also present a quantitative comparison of the proposed reconfigurable RSP with state-of-the-art RSP implementations on FPGA, as shown in Table \ref{tab:prior_art}. The table compares the hardware complexity (resources, latency, and power) alongside RSP resolution and processing dimensions. Amongst contemporary works, only the proposed work offers 3D localization in range, azimuth, and Doppler, and supports ISAC.}

\color{black}

\section{System Level Performance Analysis}
\label{sec:results}
\color{red}
\begin{table*}[!ht]
\centering
\caption{\textcolor{black}{Communication Throughput comparison for ISAC under different noise floor levels with 10 MUs.}}
 \vspace{-0.2cm}
\label{tab:ISAC_throughput_10}
\LARGE
 \resizebox{\textwidth}{!}{
 {\color{black}
 \begin{tabular}{|cc|c|c|c|c|c|c|c|c|}
\hline
\multicolumn{2}{|c|}{\multirow{4}{*}{\textbf{ISAC/Standard 802.11ad}}} &
  \multicolumn{2}{c|}{\textbf{Noise Floor: -100 dBm}} &
  \multicolumn{2}{c|}{\textbf{Noise Floor: -90 dBm}} &
  \multicolumn{2}{c|}{\textbf{Noise Floor: -80 dBm}} &
  \multicolumn{2}{c|}{\textbf{Noise Floor: -73 dBm}} \\ \cline{3-10} & &  \multicolumn{1}{c|}{\multirow{3}{*}{\textbf{\begin{tabular}[c]{@{}c@{}}Average \\Throughput \\ (Msps)\end{tabular}}}}
  &
  \multirow{3}{*}{\textbf{\begin{tabular}[c]{@{}c@{}}Beam \\  Alignments\end{tabular}}} &
  \multicolumn{1}{c|}{\multirow{3}{*}{\textbf{\begin{tabular}[c]{@{}c@{}}Average \\Throughput\\ (Msps)\end{tabular}}}}
  &
  \multirow{3}{*}{\textbf{\begin{tabular}[c]{@{}c@{}}Beam \\ Alignments\end{tabular}}} &
  \multicolumn{1}{c|}{\multirow{3}{*}{\textbf{\begin{tabular}[c]{@{}c@{}}Average\\ Throughput\\ (Msps)\end{tabular}}}} &
  \multirow{3}{*}{\textbf{\begin{tabular}[c]{@{}c@{}}Beam \\ Alignments\end{tabular}}} &
  \multicolumn{1}{c|}{  \multirow{3}{*}{\textbf{\begin{tabular}[c]{@{}c@{}}Average\\ Throughput\\ (Msps)\end{tabular}}}} &
  \multirow{3}{*}{\textbf{\begin{tabular}[c]{@{}c@{}}Beam \\ Alignments\end{tabular}}}
  \\ &  \multicolumn{1}{c|}{} &  \multicolumn{1}{c|}{}&  \multicolumn{1}{c|}{}&  \multicolumn{1}{c|}{}&  \multicolumn{1}{c|}{}&  \multicolumn{1}{c|}{}&  \multicolumn{1}{c|}{}&  \multicolumn{1}{c|}{}&  \multicolumn{1}{c|}{}\\
  &  \multicolumn{1}{c|}{} &  \multicolumn{1}{c|}{}&  \multicolumn{1}{c|}{}&  \multicolumn{1}{c|}{}&  \multicolumn{1}{c|}{}&  \multicolumn{1}{c|}{}&  \multicolumn{1}{c|}{}&  \multicolumn{1}{c|}{}&  \multicolumn{1}{c|}{}\\\hline \multicolumn{2}{|c|}{\textbf{Standard 802.11ad}}  &
  81.48  (-62.3\%)  &
  9 &
  62.16 (-70\%) &
 9  &
 21.4 (-88.7\%) &
 12 &
 21.4 (-87\%)&
 14  \\ \hline
  \multicolumn{2}{|c|}{\textbf{SARP RSP based ISAC}}  &
  \multicolumn{1}{c|}{216.31} &
  11 &
  145.93 (-29.4\%) &
  16 &
 152 (-19.6\%)  &
 19  &
 72.3 (-56\%) &
 25  \\ \hline
  
  \multicolumn{2}{|c|}{\textbf{MJARP RSP based ISAC}} & 
  173.45 (-19.8\%)&
  9 &
 174.08 (-15.8\%)&
 9  &
164 (-13.3\%) &
 10  &
 164.11 &
 10 \\ \hline
\multirow{4}{*}{\textbf{\begin{tabular}[c]{@{}c@{}}Reconfigurable\\ RSP based ISAC \end{tabular}}} & \multicolumn{1}{|c|}{\textbf{All targets:SARP}} &
  \multicolumn{1}{c|}{\textbf{216.31}} &
  \textbf{11} &
  \multicolumn{1}{c|}{145.93}  &
  16 &
  \multicolumn{1}{c|}{152} &
  19 &
  \multicolumn{1}{c|}{72.3} &
 25  \\ \cline{2-10} & 
\multicolumn{1}{|c|}{$\mathbf{T_1-T_6:SARP;T_7-T_{10}:MJARP}$} &
  \multicolumn{1}{c|}{177.6} &
  11 &
  \multicolumn{1}{c|}{\textbf{206.67}} &
  \textbf{11} &
  \multicolumn{1}{c|}{169.31} &
  14 &
  \multicolumn{1}{c|}{110.19} &
  17 \\ \cline{2-10} &
\multicolumn{1}{|c|}{$\mathbf{T_1-T_3:SARP;T_4-T_{10}:MJARP}$} &
  \multicolumn{1}{c|}{208.4} &
 9  &
  \multicolumn{1}{c|}{203.53} &
  9 &
  \multicolumn{1}{c|}{\textbf{189.16}} &
  \textbf{10} &
  \multicolumn{1}{c|}{157} &
  13 \\ \cline{2-10} &
\multicolumn{1}{|c|}{\textbf{All targets:MJARP}} &
  \multicolumn{1}{c|}{173.45} &
  9 &
  \multicolumn{1}{c|}{174.08} &
  9 &
  \multicolumn{1}{c|}{164} &
 10  &
  \multicolumn{1}{c|}{\textbf{164.11}} &
  \textbf{10} \\ \hline
\end{tabular}}
}
 \vspace{-0.2cm}
\end{table*}
\color{black}
We compare the system-level performance of ISAC, where radar sensing is used to localize the MU, identify the beam, and establish high-throughput directional communication between the BS and MU over the selected beam. We consider an ISAC BS located at the origin (0,0,0) and 10 MUs/radar targets, T1–T10 (MU1–MU10), modelled as point targets with RCS sampled from an exponential random distribution with $\sigma_{avg}=10,5,1,m^2$ for targets T1–T3, T4–T6, and T7–T10, respectively. These targets travel along separate paths with horizontal or vertical trajectories w.r.t. BS, over a duration of 1 second, with initial positions at t=0 indicated in Fig.~\ref{fig:mu_pos}. Here, out of the 10 targets, we consider 8 MUs and 2 static clutters. These static clutter scatterers are expected to be identified after RSP localization and are not served by the BS during the communication cycle. The MUs and BS have a ULA of 4 and 32 antennas, respectively. During beam alignment, we assume that alignment at MU takes place through standard training; however, since the MU processes a small number of beams, the overall alignment duration is determined by the RSP time at the BS. The Rician channel is modeled as discussed in Section II, with $\ricK=2$ dB and multipath strength of -98 dBm. 
In the ISAC system, the switch from sensing to communication happens after completion of RSP, while the switch from communication to sensing happens when SCNR at any MU degrades below a threshold. The system that quickly identifies an accurate beam (lower RSP time) and minimizes switchings (longer communication stage) will offer higher throughput and should be preferred.\\
\indent Table~7 compares the average communication throughput per MU over 1 second using the different ISAC configurations ((1) all targets:SARP, (2) 33.3\% targets:SARP, 66.6\% targets:MJARP, (3) 66.6\% targets:MJARP, 33.3\% targets: SARP, (4) all targets:MJARP) at the BS, benchmarked against standard 802.11ad for different noise floor levels. All four RSP-based ISAC approaches outperform the standard 802.11 by 62\%-88\% higher throughput across all noise floors because of the lengthy beam training procedure of the standard, which results in less time for communication. Based on the noise conditions at the BS, the optimal reconfigurable RSP configuration can be selected for best communication performance and minimal hardware complexity, as discussed in Section IV. For a noise floor of -100 dBm, processing of all targets with SARP provides the highest throughput. This is because under high SNRs, SARP offers similar performance to MJARP, leading to a comparable number of beam alignments (indicated in Table 7) and each of lesser duration due to the low processing time of SARP. While MJARP processing offers an even lower number of beam alignments, its throughput is lower than SARP due to higher processing latency. Thus, the reconfigurable RSP is best suited to switch to the (all target: SARP) configuration for this noise floor. With a further increase in noise, the reconfigurable RSP can select an alternative appropriate configuration to deliver the best communication throughput, as shown in Table 7. For noise floor -90 dBm and -80 dBm, both SARP and MJARP RSP-based ISAC offer lower throughput than the proposed reconfigurable RSP, indicating that just switching between RSP is not sufficient. Instead, the target noise floor-based switching adapted in the proposed approach is superior. For different noise levels, overall improvement in ISAC with reconfigurable RSP is substantial and varies between 13\% and 56\%.  
\textcolor{black}{The ISAC throughput is most sensitive to the RSP latency at the BS, which can vary with the number of targets in the environment, followed by the variation in the Rician factor. However, even under these variations, the proposed ISAC framework consistently outperforms prior ISAC work \cite{tewari2024reconfigrsp} and the standard 802.11ad.}

\begin{figure}[!h]
    \centering
    \includegraphics[scale = 0.4]{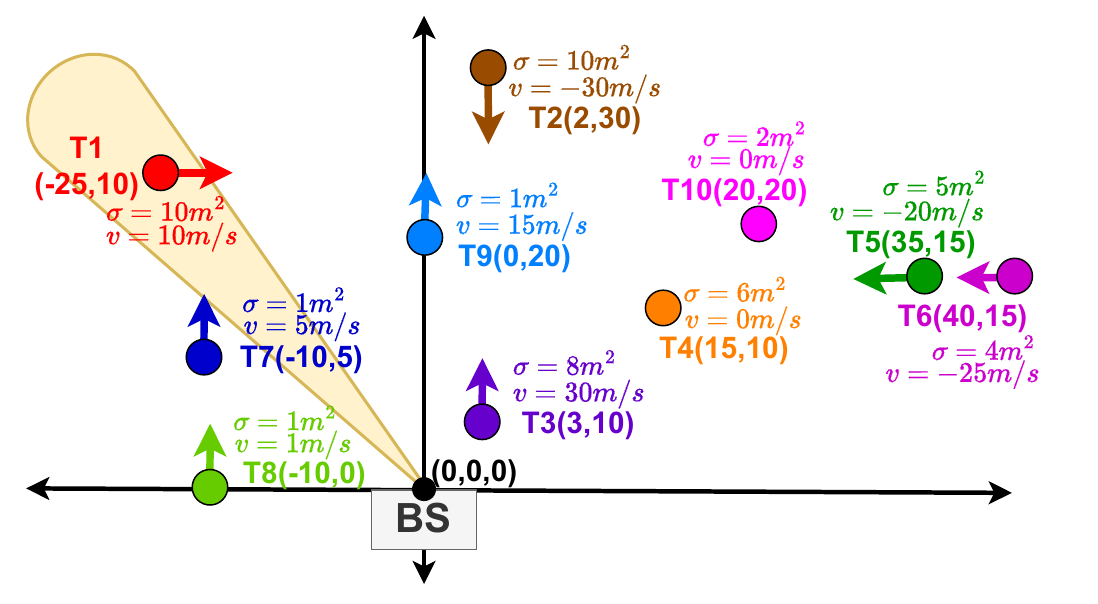}
    \vspace{-0.2cm}
\caption{\textcolor{black}{Initial positions of the 10 MUs at $t=0s$.}}
    \label{fig:mu_pos}
     \vspace{-0.25cm}
\end{figure}
\color{black}
\section{Conclusion}
\label{Sec:conc}
We presented a novel reconfigurable RSP architecture for the ISAC system offering high-speed range, azimuth, and Doppler localization of multiple targets. The reconfigurable architecture is implemented on the Zynq MPSoC platform and is optimized for low latency and resource efficiency through fixed-point analysis and reuse of hardware blocks across different RSP configurations. 
Under low noise floors, reconfigurable architecture can dynamically switch to the lower hardware complexity (latency and power consumption) SARP RSP configuration without any degradation in the RSP performance compared to the high accuracy and hardware complexity MJARP configuration. On the other hand, MJARP is the preferred configuration under a high noise floor.
The reconfigurable RSP offers up to 5 times reduction in RSP time and 26.4\% reduction in dynamic power consumption over state-of-the-art for the localization of three targets.
The reconfigurable RSP with faster beam alignment provides up to 24\% improvement in communication throughput of ISAC over the MJARP RSP and 88\% over standard 802.11ad. Future work will focus on performance analysis of reconfigurable RSP for different waveforms and extended radar target models.

\section*{Appendix}
\subsection{System Level Performance Analysis with Three Mobile Users}
This section analyzes the RSP performance and studies the impact of radar metrics on communications in the ISAC setup. We consider an ISAC BS located at the origin (0,0,0) and three mobile users/radar targets, MU1, MU2, and MU3, modelled as point targets with average RCS, $\sigma=10,5,1 m^2$ respectively. These targets travel along separate paths over a duration of 1 second, as shown in Figure \ref{fig:comm_ber}. (a). MU1 and MU2 move along a transverse trajectory, whereas MU3 follows a radial trajectory with respect to the BS. The MU and BS have a ULA of 4 and 32 antennas, respectively. The Rician channel is modeled with $\ricK=2 dB$, and multipath strength of -98 dBm.
\begin{figure*}[h]
    \centering
    \includegraphics[scale = 0.55]{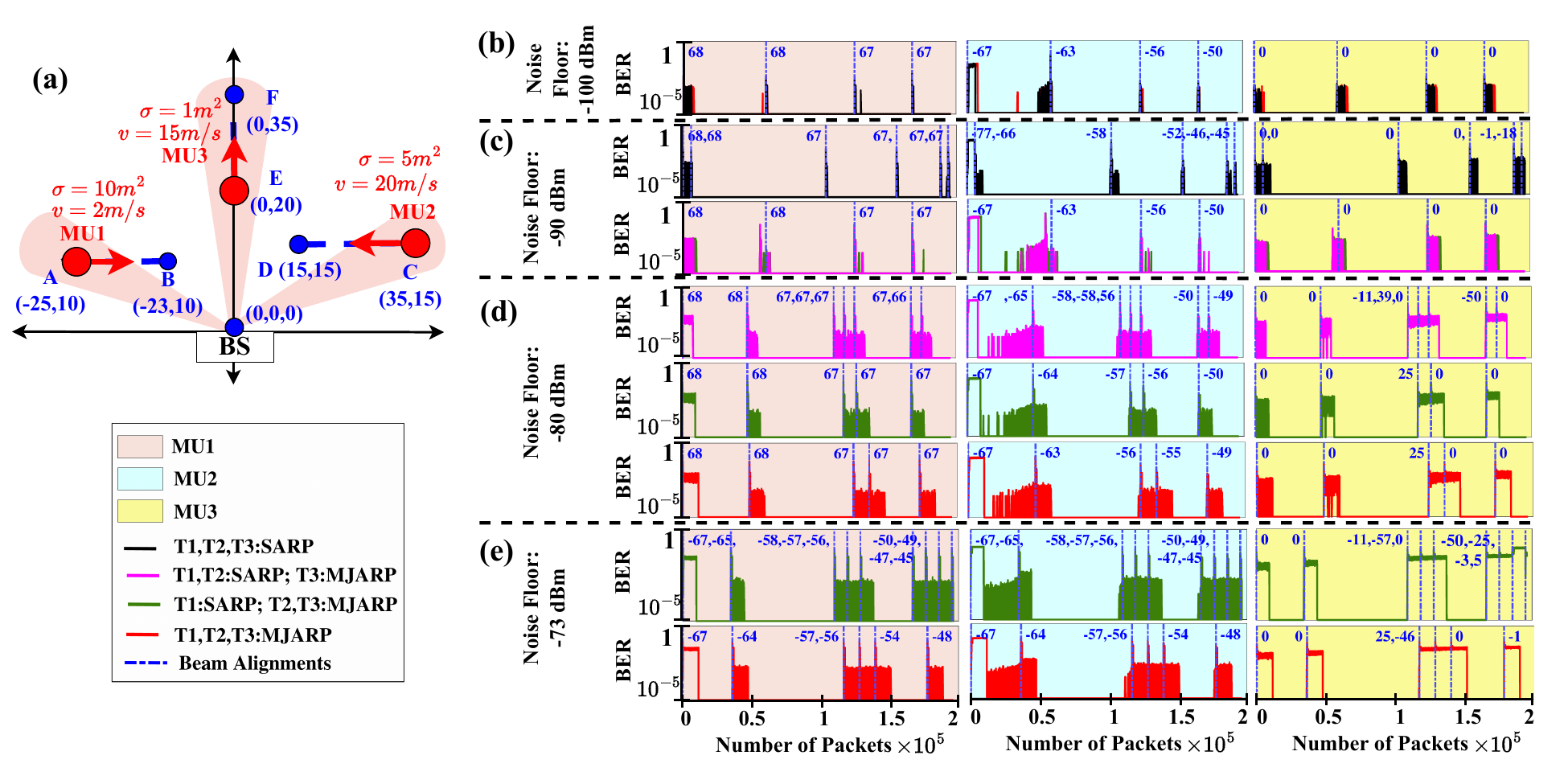}
\caption{(a)Trajectory of the three MUs over 1 second, bit-error-rate over 1 second under noise floor (b)-100 dBm, (c)-90 dBm, (d)-80 dBm, and (e)-73 dBm. }
    \label{fig:comm_ber}
\end{figure*}
\begin{table*}[b]
\centering
\caption{Communication Throughput comparison between standard 802.11ad and ISAC with JARP and reconfigurable RSP for different noise floors.
}
\label{tab:ISAC_throughput}
\LARGE
 \resizebox{\textwidth}{!}{
 \begin{tabular}{|cc|cc|cc|cc|cc|}
\hline
\multicolumn{2}{|c|}{\multirow{4}{*}{\textbf{ISAC/Standard 802.11ad}}} &
  \multicolumn{2}{c|}{\textbf{Noise Floor: -100 dBm}} &
  \multicolumn{2}{c|}{\textbf{Noise Floor: -90 dBm}} &
  \multicolumn{2}{c|}{\textbf{Noise Floor: -80 dBm}} &
  \multicolumn{2}{c|}{\textbf{Noise Floor: -73 dBm}} \\ \cline{3-10} & &  \multicolumn{1}{c|}{\multirow{3}{*}{\textbf{\begin{tabular}[c]{@{}c@{}}Average \\Throughput \\ (Msps)\end{tabular}}}}
  &
  \multirow{3}{*}{\textbf{\begin{tabular}[c]{@{}c@{}}Beam \\  Alignments\end{tabular}}} &
  \multicolumn{1}{c|}{\multirow{3}{*}{\textbf{\begin{tabular}[c]{@{}c@{}}Average \\Throughput\\ (Msps)\end{tabular}}}}
  &
  \multirow{3}{*}{\textbf{\begin{tabular}[c]{@{}c@{}}Beam \\ Alignments\end{tabular}}} &
  \multicolumn{1}{c|}{\multirow{3}{*}{\textbf{\begin{tabular}[c]{@{}c@{}}Average\\ Throughput\\ (Msps)\end{tabular}}}} &
  \multirow{3}{*}{\textbf{\begin{tabular}[c]{@{}c@{}}Beam \\ Alignments\end{tabular}}} &
  \multicolumn{1}{c|}{  \multirow{3}{*}{\textbf{\begin{tabular}[c]{@{}c@{}}Average\\ Throughput\\ (Msps)\end{tabular}}}} &
  \multirow{3}{*}{\textbf{\begin{tabular}[c]{@{}c@{}}Beam \\ Alignments\end{tabular}}}
  \\ &  \multicolumn{1}{c|}{} &  \multicolumn{1}{c|}{}&  \multicolumn{1}{c|}{}&  \multicolumn{1}{c|}{}&  \multicolumn{1}{c|}{}&  \multicolumn{1}{c|}{}&  \multicolumn{1}{c|}{}&  \multicolumn{1}{c|}{}&  \multicolumn{1}{c|}{}\\
  &  \multicolumn{1}{c|}{} &  \multicolumn{1}{c|}{}&  \multicolumn{1}{c|}{}&  \multicolumn{1}{c|}{}&  \multicolumn{1}{c|}{}&  \multicolumn{1}{c|}{}&  \multicolumn{1}{c|}{}&  \multicolumn{1}{c|}{}&  \multicolumn{1}{c|}{}\\\hline \multicolumn{2}{|c|}{\textbf{Standard 802.11ad}}  &
  \multicolumn{1}{c|}{\begin{tabular}[c]{@{}c@{}}271.57\\  (-48.35\%) \end{tabular}} &
  4 &
  \multicolumn{1}{c|}{\begin{tabular}[c]{@{}c@{}}207.24\\ (-58.7\%) \end{tabular}} &
  8 &
  \multicolumn{1}{c|}{\begin{tabular}[c]{@{}c@{}}71.26 \\ (-84.2\%) \end{tabular}} &
  13 &
  \multicolumn{1}{c|}{\begin{tabular}[c]{@{}c@{}}71.25 \\ (-81.6\%) \end{tabular}} &
  13 \\ \hline
  \multicolumn{2}{|c|}{\textbf{SARP RSP based ISAC}}  &
  \multicolumn{1}{c|}{524.79} &
  4 &
  \multicolumn{1}{c|}{\begin{tabular}[c]{@{}c@{}}471.17 \\ (-6.14\%) \end{tabular}}  &
  6 &
  \multicolumn{1}{c|}{\begin{tabular}[c]{@{}c@{}}370 \\ (-17.84\%) \end{tabular}} &
  13 &
  \multicolumn{1}{c|}{\begin{tabular}[c]{@{}c@{}}335.55 \\ (-15.14\%) \end{tabular}} &
  15 \\ \hline
  
  \multicolumn{2}{|c|}{\textbf{MJARP RSP based ISAC}} &
  \multicolumn{1}{c|}{\begin{tabular}[c]{@{}c@{}}455.93 \\ (-13.12\%) \end{tabular}} &
  4 &
  \multicolumn{1}{c|}{\begin{tabular}[c]{@{}c@{}}455.97 \\ (-9.2\%) \end{tabular}} &
  4 &
  \multicolumn{1}{c|}{\begin{tabular}[c]{@{}c@{}}421.54 \\ (-6.4\%) \end{tabular}} &
  5 &
  \multicolumn{1}{c|}{387.12} &
  6 \\ \hline
\multirow{4}{*}{\textbf{\begin{tabular}[c]{@{}c@{}}Reconfigurable\\ RSP based ISAC \end{tabular}}} & \multicolumn{1}{|c|}{$\mathbf{T_1,T_2,T_3:SARP}$} &
  \multicolumn{1}{c|}{\textbf{524.79}} &
  \textbf{4} &
  \multicolumn{1}{c|}{471.17}  &
  6 &
  \multicolumn{1}{c|}{370} &
  13 &
  \multicolumn{1}{c|}{335.55} &
  15 \\ \cline{2-10} & 
\multicolumn{1}{|c|}{$\mathbf{T_1,T_2:SARP;T_3:MJARP}$} &
  \multicolumn{1}{c|}{502.04} &
  4 &
  \multicolumn{1}{c|}{\textbf{502}} &
  \textbf{4} &
  \multicolumn{1}{c|}{433.33} &
  7 &
  \multicolumn{1}{c|}{295.08} &
  12 \\ \cline{2-10} &
\multicolumn{1}{|c|}{$\mathbf{T_1:SARP;T_2,T_3:MJARP}$} &
  \multicolumn{1}{c|}{479} &
  4 &
  \multicolumn{1}{c|}{478.96} &
  4 &
  \multicolumn{1}{c|}{\textbf{450.34}} &
  \textbf{5} &
  \multicolumn{1}{c|}{284.57} &
  9 \\ \cline{2-10} &
\multicolumn{1}{|c|}{$\mathbf{T_1,T_2,T_3:MJARP}$} &
  \multicolumn{1}{c|}{455.93} &
  4 &
  \multicolumn{1}{c|}{455.97} &
  4 &
  \multicolumn{1}{c|}{421.54} &
  5 &
  \multicolumn{1}{c|}{\textbf{387.12}} &
  \textbf{6} \\ \hline
\end{tabular}
}
\end{table*}
At the start (time zero seconds), the beam alignment stage commences between the BS and the MUs. During this stage, the BS transmits packets omnidirectionally and processes the target-reflected echoes via RSP to localize multiple MUs in the environment. Each MU finds its best beam towards BS by performing standard 802.11ad processing on the received downlink packets. We assume that the beam search time taken by MU is always shorter than the RSP time at BS due to a smaller ULA size, leading to fewer candidate search beams at MU. Hence, the beam alignment duration is modeled based on the RSP execution time. The beam alignment stage is followed by the communication stage, where the BS and MUs know the best beam towards each other, and communicate over highly directional links using analog beamforming. Here, the BS serves the MUs in a time division multiplexing manner. The BS switches back to the beam alignment stage whenever the SCNR at the BS, for any one of the MUs, drops below a certain threshold. The bit error rate (BER) is evaluated at each MU, and remains high during the beam alignment phase and drops during the communication phase.

We evaluate the communication link metrics in terms of BER and throughput for each MU under different noise floor levels. Table \ref{tab:ISAC_throughput} presents the link throughput calculated during the communication stage over 1 second using the different proposed ISAC configurations at the BS, benchmarked with standard 802.11ad processing.  The ISAC of all RSP configurations outperforms the standard 802.11 under all noise floors by avoiding the lengthy beam-alignment stage in the standard, which involves the time-consuming beam-training procedure between BS and MUs. Based on the noise conditions at the BS, the optimal reconfigurable RSP architecture for ISAC, out of the four possible configurations discussed in Section. IV of the main manuscript can be selected for the best communication performance and lowest hardware complexity. Figure \ref{fig:comm_ber}. (b) shows the BER of the three MUs, under a noise floor of -100 dBm, with SARP and MJARP RSP for all three targets at the BS. Due to high system SCNR, the RSP performance of SARP is identical to MJARP, leading to the same number of beam alignment stages for both over 1 second. In Figure \ref{fig:comm_ber}, the time instants indicating the start of the beam alignment stages are marked with a blue dotted line along with the estimated azimuth of the MU computed from RSP at BS during beam alignment. In Figure \ref{fig:comm_ber}. (b), the 4 beam alignments occur at the same time instant for all three targets. However, since the RSP latency of MJARP is higher than SARP, as mentioned in Table \ref{tab:rsp_hwcmp}, it leads to a longer beam alignment duration in the case of MJARP over SARP. Here, SARP RSP shows higher throughput than MJARP, becoming the preferred configuration for this case. Next, for a noise floor of -90 dBm, the number of beam alignments increases to 6 when all targets are processed with SARP RSP. This is due to incorrect localization of the third (weakest strength) target, leading to misalignments between MU3 and BS, which initiates more beam alignment stages and causes a drop in throughput as shown in Table \ref{tab:ISAC_throughput}. Switching to MJARP RSP for the $T3$, keeping the $T1$ and $T2$ in SARP, reduces the beam alignments to 4, hence improving the throughput. This makes ($T_1$,$T_2$:SARP; $T_3$:MJARP) the preferred configuration for this case. As shown in Table \ref{tab:ISAC_throughput}, keeping $T_1$ in SARP and $T_2$ and $T_3$ in MJARP also provides 4 beam alignments; however, the throughput is lower due to the longer duration for establishing beam alignment. Similarly, as can be observed from Figure \ref{fig:comm_ber}. (d,e) and Table \ref{tab:ISAC_throughput}, ($T_1$:SARP; $T_2$,$T_3$:MJARP) and ($T_1$,$T_2$,$T_3$:MJARP) are the most suitable RSP configurations under higher noise floor of -80 dBm and -73 dBm respectively. Thus, based upon the noise floor switch points, the reconfigurable RSP can select the appropriate configuration for the most optimum ISAC performance across all MUs.
\subsection{Dynamic Power Breakdown}
The reconfigurable RSP architecture has a total power consumption of 4.5-4.7 $W$ with the detailed power breakdown as mentioned in the following Table \ref{tab:power_breakdown}. These results are derived from the Xilinx-Vivado post-implementation power utilization report. The total dynamic power consumption on Zynq MPSoC is 2.43-3.43 $W$ as mentioned in {Table VIII, Section V.C, page 12}. Table \ref{tab:power_breakdown} provides the detailed dynamic power breakdown of all the major blocks presented in Figure \ref{fig:rsp_hw}, {Section III.C, page 6}. Here, the FPGA clock frequency is 100 $MHz$. The SARP and MJARP configurations of reconfigurable RSP differ in their dynamic power consumption due to the selective activation of the MF, DBF, and PS blocks by the FSM shown in Figure \ref{fig:rsp_hw}, based on the selected RSP configuration. The remaining blocks, such as Selective MF and MUSIC, are common to both configurations. As expected, the dynamic power consumption of SARP is lower than that of MJARP. The total static power for the complete design on the Zynq MPSoC is reported as shown.

The energy per frame for the localization of three targets with SARP RSP is 0.13$J$, and that of MJARP is 0.27$J$ at an FPGA clock of 100 $MHz$.

\begin{table}[!h]
\caption{Static and Dynamic Power of different RSP blocks at clock frequency of 100 $MHz$  }
\label{tab:power_breakdown}
\huge
\centering
\renewcommand{\arraystretch}{1}
\resizebox{\linewidth}{!}{
\begin{tabular}{|cc|ccc|}
\hline
\multicolumn{1}{|c|}{\multirow{2}{*}{\textbf{Configuration}}} &
  \multirow{2}{*}{\textbf{RSP Block}} &
  \multicolumn{2}{c|}{\textbf{Dynamic   Power}} &
  \multirow{2}{*}{\textbf{\begin{tabular}[c]{@{}c@{}}Static   Power   \\  (W)\end{tabular}}} \\ \cline{3-4}
\multicolumn{1}{|c|}{}                            &                     & \multicolumn{1}{c|}{\textbf{Watts}} & \multicolumn{1}{c|}{\textbf{\% w.r.t.   total}} &  \\ \hline
\multicolumn{1}{|c|}{\multirow{3}{*}{SARP}} &
  MF &
  \multicolumn{1}{c|}{0.092} &
  \multicolumn{1}{c|}{2\%} &
  \multirow{11}{*}{\begin{tabular}[c]{@{}c@{}}1.22   \\ (25.8\%)\end{tabular}} \\ \cline{2-4}
\multicolumn{1}{|c|}{}                            & DBF                 & \multicolumn{1}{c|}{0.115}          & \multicolumn{1}{c|}{2.6\%}                      &  \\ \cline{2-4}
\multicolumn{1}{|c|}{}                            & CLEAN + peak search & \multicolumn{1}{c|}{0.142}          & \multicolumn{1}{c|}{3.1\%}                      &  \\ \cline{1-4}
\multicolumn{1}{|c|}{\multirow{3}{*}{MJARP}}      & MF                  & \multicolumn{1}{c|}{0.116}          & \multicolumn{1}{c|}{2.5\%}                      &  \\ \cline{2-4}
\multicolumn{1}{|c|}{}                            & DBF                 & \multicolumn{1}{c|}{0.183}          & \multicolumn{1}{c|}{4\%}                        &  \\ \cline{2-4}
\multicolumn{1}{|c|}{}                            & CLEAN + peak search & \multicolumn{1}{c|}{0.25}           & \multicolumn{1}{c|}{5.4\%}                      &  \\ \cline{1-4}
\multicolumn{1}{|c|}{\multirow{5}{*}{RSP blocks}} & PSR                 & \multicolumn{1}{c|}{0.078}          & \multicolumn{1}{c|}{1.7\%}                      &  \\ \cline{2-4}
\multicolumn{1}{|c|}{}                            & Selective MF        & \multicolumn{1}{c|}{0.132}          & \multicolumn{1}{c|}{2.8\%}                      &  \\ \cline{2-4}
\multicolumn{1}{|c|}{}                            & MUSIC               & \multicolumn{1}{c|}{0.998}          & \multicolumn{1}{c|}{21.4\%}                     &  \\ \cline{2-4}
\multicolumn{1}{|c|}{}                            & ARM processor       & \multicolumn{1}{c|}{1.65}           & \multicolumn{1}{c|}{35.5\%}                     &  \\ \cline{2-4}
\multicolumn{1}{|c|}{}                            & DMA                 & \multicolumn{1}{c|}{0.02}           & \multicolumn{1}{c|}{9.3\%}                      &  \\ \hline
\multicolumn{2}{|c|}{\textbf{Total Power SARP (W)}}                     & \multicolumn{3}{c|}{\textbf{4.45}}                                                       \\ \hline
\multicolumn{2}{|c|}{\textbf{Total Power MJARP (W)}}                    & \multicolumn{3}{c|}{\textbf{4.65}}                                                       \\ \hline
\end{tabular}
}
\end{table}

\subsection{Sensitivity of ISAC Throughput }
\subsubsection{Effect of RSP latency at BS}
The RSP latency can vary based on the following two factors:
\begin{enumerate}
\item The selection of the appropriate RSP algorithm based on channel conditions, low latency-low accuracy SARP, or high latency-high accuracy MJARP.
\item The number of radar targets/MU to be localized in the environment.
\end{enumerate}

We present the effect of increasing the number of targets on the ISAC throughput. With the increase in targets, RSP latency also increases due to multiple CLEAN iterations for localization in the range-azimuth domain first, followed by the corresponding Doppler processing for each localization. Figure \ref{fig:target_Throughput} compares the communication throughput of the proposed ISAC system with reconfigurable RSP, ISAC presented in prior work \cite{tewari2024reconfigrsp}, and standard 802.11ad for a noise floor of -73 $dB$. Here, it can be seen that the proposed ISAC outperforms \cite{tewari2024reconfigrsp} and the standard for any number of targets. However, there is an overall decline in throughput with an increasing number of targets, due to the following reasons-
\begin{enumerate}[label=\alph*.]
\item Since radar and communication time share the ISAC infrastructure, there is an increase in RSP latency due to the localization of a large number of targets, which in turn shortens the communication cycle available for servicing the MU, leading to a reduction in throughput. 
Here, the reconfigurable RSP is more robust to target variations, as it can switch to the low-latency RSP configuration SARP to maintain optimal performance, compared to \cite{tewari2024reconfigrsp}.
\item After the beam alignment stage, the BS services the MUs in a time division multiplexing (TDM) manner. Thus, as the number of MUs increases, the service time available to each MU decreases, reducing the average throughput per MU. This effect can be minimized by using a multi-user multiple-input multiple-output (MIMO) system that can serve all MUs simultaneously during the communication stage. We plan to explore this strategy in the future.
\end{enumerate}
\begin{figure}[!h]
    \centering
\includegraphics[scale = 0.45]{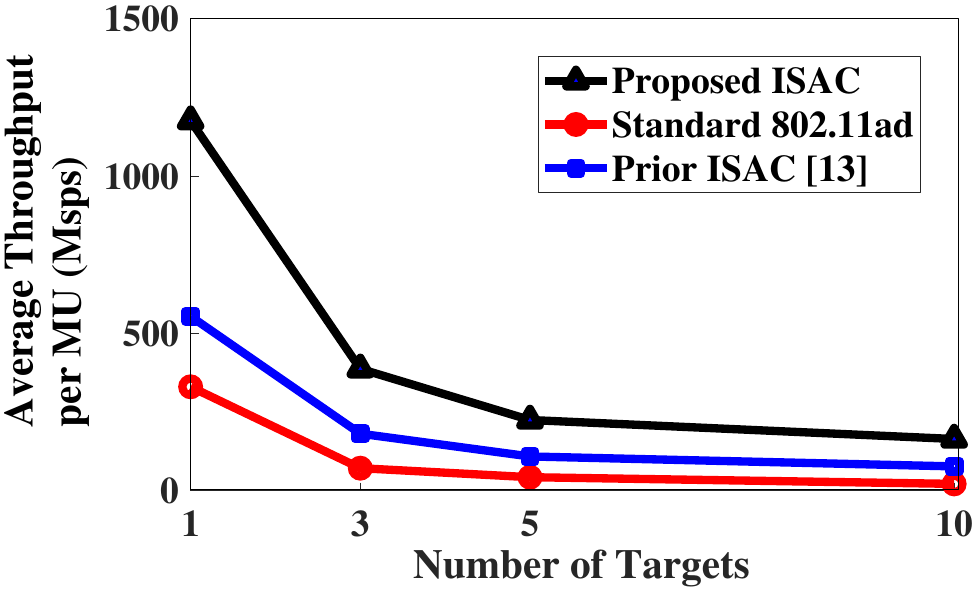}
    \caption{Variation in ISAC throughput per MU with increase in number of targets under noise floor of -73 dBm.}    \label{fig:target_Throughput}
\end{figure}

\subsubsection{Effect of Rician Factor ($\ricK$)}
The throughput results presented in {Section VI, page 13} of the revised manuscript have been evaluated for $\ricK=2 dB$. We use the same simulation setup with 10 MUs at the positions shown in Figure \ref{fig:mu_pos}, and analyze the effect of varying $\ricK$ from 0$dB$ to 15$dB$ on the ISAC throughput. The results are highlighted in Figure \ref{fig:k_Throughput}. High values of $\ricK$ correspond to a strong line-of-sight (LOS) component, whereas low values of $\ricK$ correspond to a weak LOS component in the received signal. The throughput curves show an identical trend for both noise floors of -73 $dBm$ and -100 $dBm$; there is a sharp increase from 0 $dB$ to 2 $dB$, while throughput reaches asymptotic convergence for moderate and high values of $\ricK$. The proposed ISAC outperforms \cite{tewari2024reconfigrsp} across all $\ricK$. The average throughput of standard 802.11ad across all $\ricK$ is very poor, 40 $Msps$, and hence not highlighted in Figure \ref{fig:k_Throughput}.
\begin{figure}[!h]
    \centering
\includegraphics[scale = 0.45]{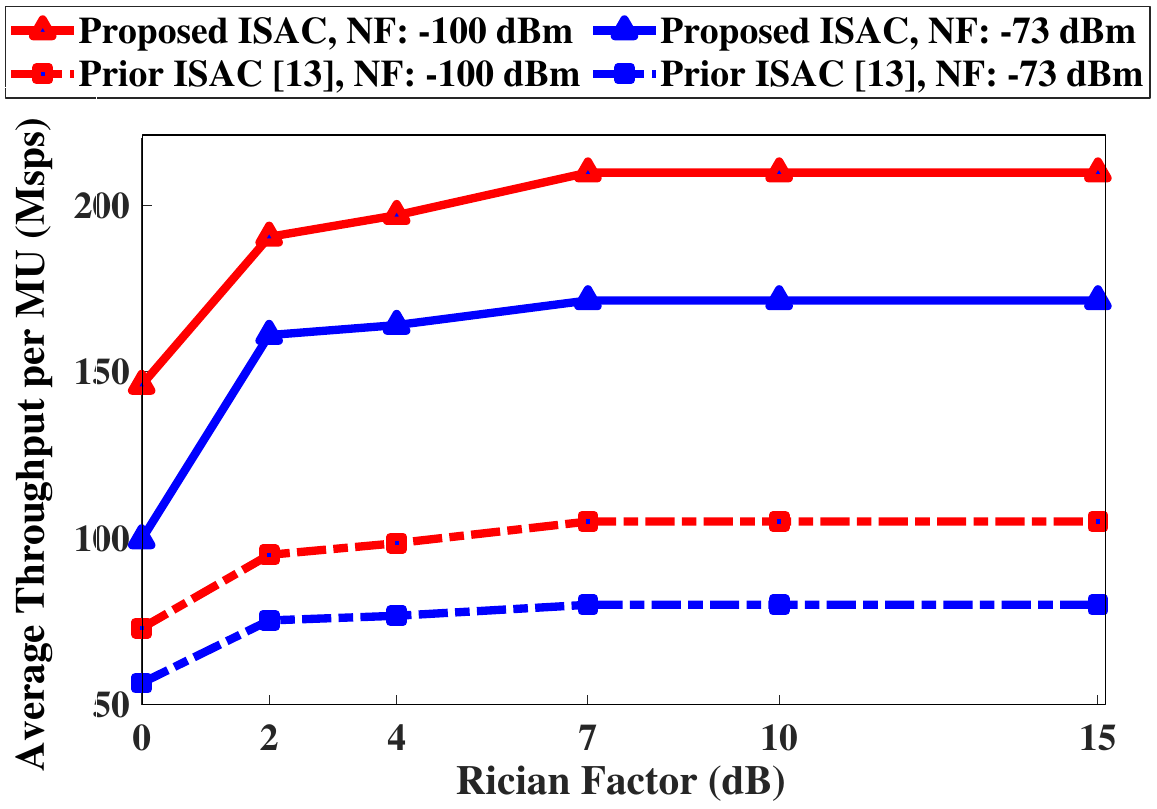}
    \caption{Variation in throughput per MU for different Rician factors. }    \label{fig:k_Throughput}
\end{figure}

\subsection{Hardware Implementation of JARP, MJARP, and reconfigurable RSP through Vivado Design Flow}
In this section, we present the detailed hardware designs of the different RSP configurations discussed in 
Table \ref{tab:hw_cmpl} (Section V.C of the main manuscript). Table \ref{tab:hw_cmpl} shows the hardware-software co-design profiling analysis for the reconfigurable RSP. Here, we present the joint azimuth-range processing (JARP), modified joint azimuth-range processing (MJARP), and reconfigurable RSP configurations. The reconfigurable RSP can on-the-fly switch between the sequential azimuth-range processing (SARP) and MJARP configurations based on the noise floor level. 

The hardware designs are developed using the Xilinx Vivado design suite. The RSP hardware IP cores are developed using the Vivado High Level Synthesis (HLS) tool. These IPs are then integrated into the Vivado block design environment for generating the FPGA bitstream. The bitstream is deployed and tested on the AMD Xilinx MPSoC platform through the PYNQ workflow. The MPSoC board hosts a PYNQ-based operating system that provides Python-based drivers for controlling the hardware accelerators. Additionally, the PYNQ framework can also host Monte-Carlo simulations for obtaining the RSP localization metrics and visualization of the results in real-time.
\subsubsection{Range, azimuth, and Doppler estimation with JARP RSP}
In JARP configuration, as shown in Table \ref{tab:hw_cmpl}, MF and DBF for range-azimuth (RA) processing and MUSIC for Doppler processing are carried out in PL, whereas peak search and CLEAN are conducted in PS. The Vivado block design for the same is presented in Figure \ref{fig:jarp_rsp}. This design is for the processing of 32 packets. The PS provides the control information to the Direct Memory Access (DMA) and hardware IP cores via the AXI-Lite interface, as highlighted in green color. First, 32 $(P\times L)$ packets are sent from PS as input to the JARP accelerator in PL via DMA-1, as shown in red color. After MF and DBF, 32 $(P \times I)$ packets are returned back to PS for RA peak search and CLEAN. However, since the MF and DBF in PSR generation for CLEAN are conducted in PL itself, there are two additional packet transfers for localizing three targets. Once all three targets are localized, the slow time vectors for each target are created in PS and sent one-by-one to the MUSIC block in PL via DMA-2, shown in magenta color. The MUSIC output spectrum is then sent back to PL for Doppler velocity peak search. The JARP processing is slow due to large data transfers between PS and PL and the complete RA processing for $I$ search angles for all 32 packets.
\begin{figure*}[!t]
    \centering
    \includegraphics[scale = 0.28]{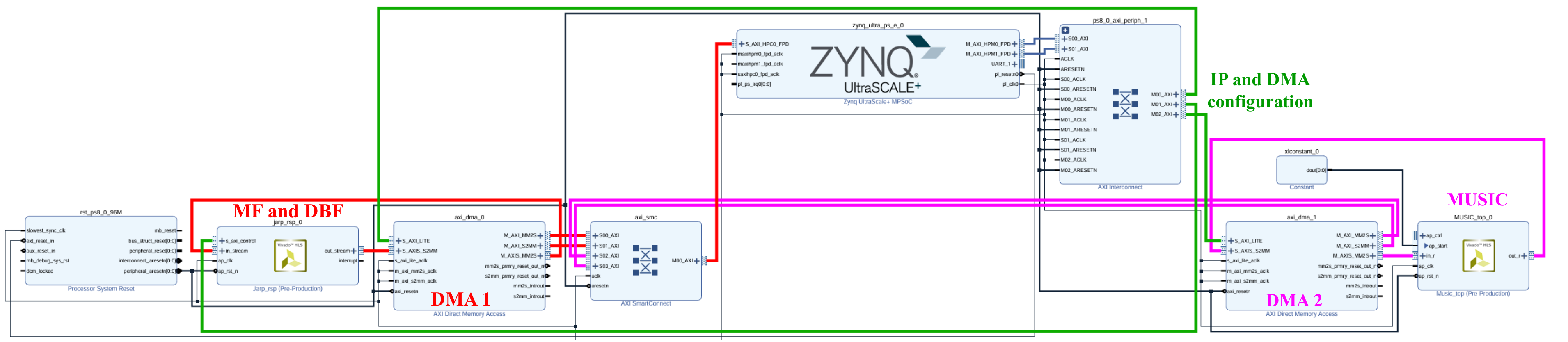}
\caption{Vivado block design for JARP architecture for range-azimuth, and Doppler estimation.  }
    \label{fig:jarp_rsp}
\end{figure*}
\subsubsection{Range and azimuth estimation with MJARP RSP}
The MJARP configuration for RA localization on the first packet is shown in the fourth row of Table \ref{tab:hw_cmpl} with optimization (a). Figure \ref{fig:opt_a} shows the Vivado block design for the same. RA peak search and CLEAN are shifted to the PL accelerator along with MF and DBF, avoiding the additional data transfers and speeding up the RA localization. Now, only the RA detections of the three targets are sent back to PS via DMA instead of the complete $(P \times I)$ RA ambiguity spectrum.
\begin{figure*}[!t]
    \centering
    \includegraphics[scale = 0.28]{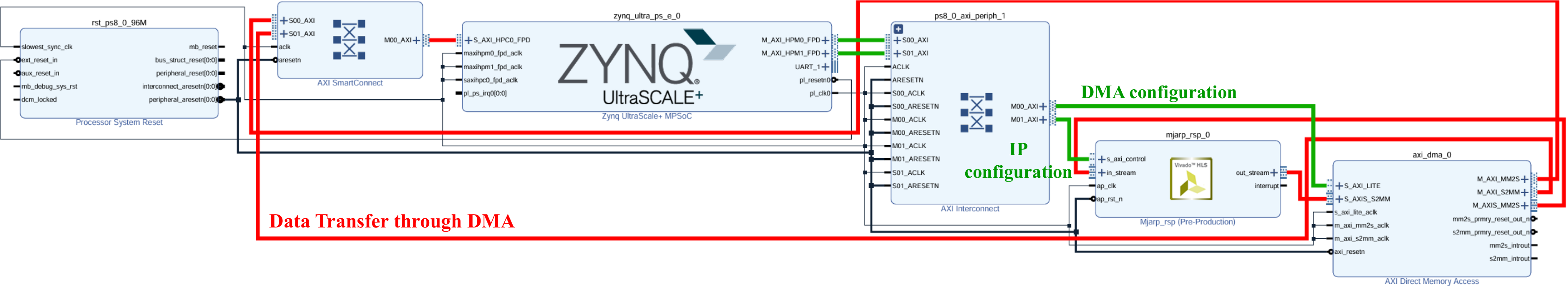}
\caption{Vivado block design for MJARP architecture for range-azimuth estimation showing optimization (a) in Table \ref{tab:hw_cmpl}.}
    \label{fig:opt_a}
\end{figure*}
\subsubsection{Range and azimuth estimation with Reconfigurable RSP}
The reconfigurable RA processing for single and 32 packets is shown in the fifth and seventh rows of Table \ref{tab:hw_cmpl} and indicated with optimization (b) and (c), respectively. Figure \ref{fig:opt_b_c} shows the Vivado block design for the same. Here, the RSP accelerator can be reconfigured to switch between the SARP and MJARP RSP approaches, and different number of packets. The configuration information for the same is provided by PS via AXI-Lite port as highlighted in green in Figure \ref{fig:opt_b_c}. The reconfigurable RSP further accelerates the RA processing compared to MJARP. In optimization (b), which is for the processing of a single packet, the speedup is due to SARP processing, with the maximum acceleration seen when all three targets are localized using SARP. On the other hand, even higher speedup occurs in optimization (c), due to SARP processing on the first packet and selective RA processing of the remaining packets. 
 \begin{figure*}[!t]
    \centering
    \includegraphics[scale = 0.28]{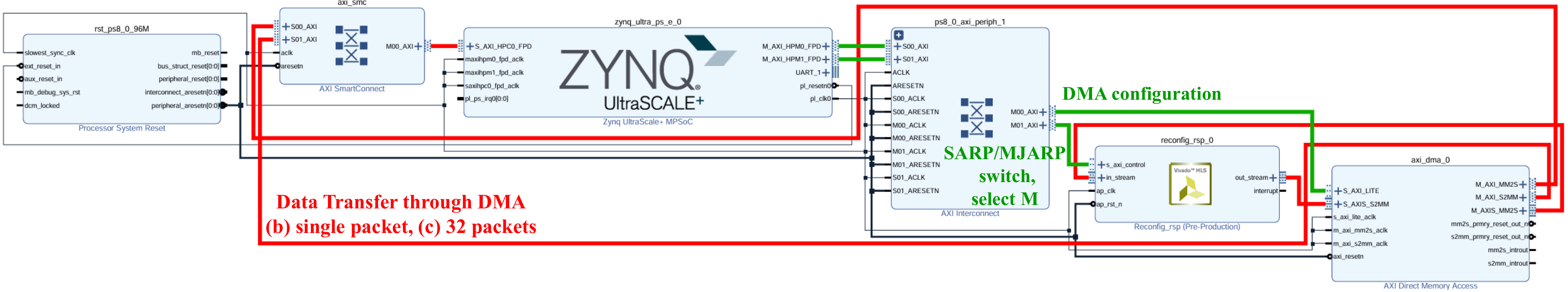}
\caption{Vivado block design for reconfigurable architecture for range-azimuth estimation showing optimization (b) and (c) in Table \ref{tab:hw_cmpl}.  }
    \label{fig:opt_b_c}
\end{figure*}
\subsubsection{Range, azimuth, and Doppler estimation with Reconfigurable RSP}
In this configuration, the RSP for 3D localization of multiple targets is completely offloaded to PL. This is shown in the last row of Table \ref{tab:hw_cmpl}, by optimization (d). As shown in Figure \ref{fig:opt_d}, the RSP accelerator is configured by PS only once at the start. The 32 packets are sent via DMA as input to the accelerator, and after processing, the range, azimuth, and Doppler indices of the three targets are transferred back to PS via DMA. Unlike the JARP implementation shown in Figure \ref{fig:jarp_rsp}, only a single DMA is used since the slow-time vector creation for the three targets takes place in PL itself after RA processing. These slow-time vectors are directly sent to MUSIC. The MUSIC accelerator is implemented in fixed-point and provides high acceleration compared to JARP. Figure \ref{fig:hw_output} shows the output range-azimuth, and Doppler spectra and localization estimates from reconfigurable RSP.
 \begin{figure*}[h]
    \centering
    \includegraphics[scale = 0.28]{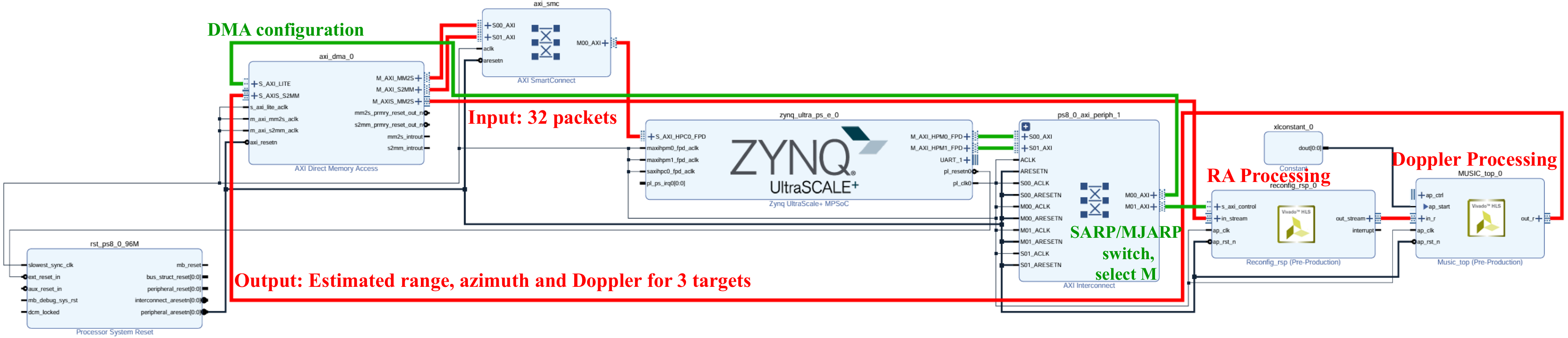}
\caption{Vivado block design for reconfigurable architecture for range, azimuth, and Doppler estimation showing optimization (d) in Table \ref{tab:hw_cmpl}.  }
    \label{fig:opt_d}
\end{figure*}

 \begin{figure*}[!h]
    \centering
    \includegraphics[scale = 0.45]{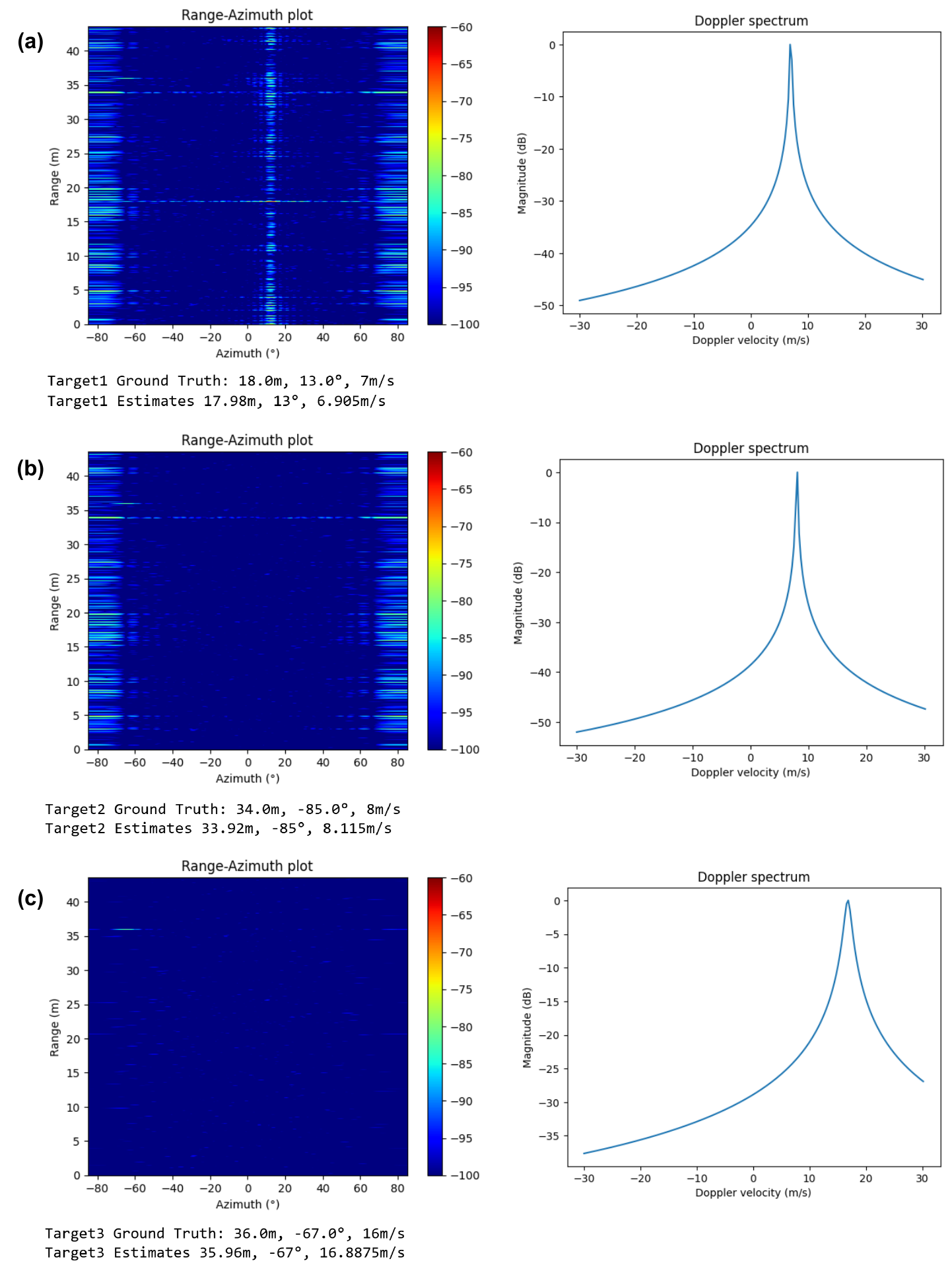}
\caption{Range-azimuth ambiguity plots and MUSIC spectrum along with localization estimates for (a) target 1, (b) target 2, and (c) target 3  after 3D processing with reconfigurable RSP on Zynq MPSoC.}
    \label{fig:hw_output}
\end{figure*}


\normalem
\bibliographystyle{ieeetr}
\bibliography{references}

@ARTICLE{tewari2024reconfigrsp,
  author={Tewari, Aakanksha and Jha, Shragvi Sidharth and Sneh, Akanksha and Darak, Sumit J and Ram, Shobha Sundar},
  journal={IEEE Transactions on Aerospace and Electronic Systems}, 
  title={Reconfigurable Radar Signal Processing Accelerator for Integrated Sensing and Communication System}, 
  year={2024},
  volume={},
  number={},
  pages={1-19},
  doi={10.1109/TAES.2024.3446462}}

@article{ram2022optimization,
  title={Optimization of Network Throughput of Joint Radar Communication System Using Stochastic Geometry},
  author={Ram, Shobha Sundar and Singhal, Shubhi and Ghatak, Gourab},
  journal={Frontiers in Signal Processing},
  volume={2},
  year={2022},
  publisher={Frontiers}
}

@article{noauthor_ieee_2016-1,
	title = {{IEEE} {Std} 802.11™-2016, {IEEE} Standard for Information technology—Telecommunications and information exchange between systems—Local and metropolitan area networks—Specific requirements—{Part} 11: {Wireless} {LAN} Medium Access Control},
	language = {en},
	year = {2016},
	pages = {3534},
	
}

@ARTICLE{sneh2022ieee,
  author={Sneh, Akanksha and Jain, Soumya and Sindhu, V Sri and Ram, Shobha Sundar and Darak, Sumit},
  journal={IEEE Transactions on Vehicular Technology}, 
  title={IEEE 802.11ad Based Joint Radar Communication Transceiver: Design, Prototype and Performance Analysis}, 
  year={2023},
  volume={},
  number={},
  pages={1-16},
  doi={10.1109/TVT.2023.3333009}}

@article{duggal2020doppler,
  title={Doppler-resilient 802.11 ad-based ultrashort range automotive joint radar-communications system},
  author={Duggal, Gaurav and Vishwakarma, Shelly and Mishra, Kumar Vijay and Ram, Shobha Sundar},
  journal={IEEE Transactions on Aerospace and Electronic Systems},
  volume={56},
  number={5},
  pages={4035--4048},
  year={2020},
  publisher={IEEE}
}

@ARTICLE{MAB_JSTSP,
  author={Sneh, Akanksha and Ram, Shobha Sundar and Darak, Sumit J and Tewari, Aakanksha},
  journal={IEEE Journal of Selected Topics in Signal Processing}, 
  title={Beam Alignment in Multipath Environments for Integrated Sensing and Communication Using Bandit Learning}, 
  year={2024},
  volume={},
  number={},
  pages={1-15},
  doi={10.1109/JSTSP.2024.3391905}}

@ARTICLE{9737357,
  author={Liu, Fan and Cui, Yuanhao and Masouros, Christos and Xu, Jie and Han, Tony Xiao and Eldar, Yonina C. and Buzzi, Stefano},
  journal={IEEE Journal on Selected Areas in Communications}, 
  title={Integrated Sensing and Communications: Toward Dual-Functional Wireless Networks for 6G and Beyond}, 
  year={2022},
  volume={40},
  number={6},
  pages={1728-1767},
  doi={10.1109/JSAC.2022.3156632}}

@ARTICLE{kvm2019mmwjrc,
  author={Mishra, Kumar Vijay and Bhavani Shankar, M.R. and Koivunen, Visa and Ottersten, Bjorn and Vorobyov, Sergiy A.},
  journal={IEEE Signal Processing Magazine}, 
  title={Toward Millimeter-Wave Joint Radar Communications: A Signal Processing Perspective}, 
  year={2019},
  volume={36},
  number={5},
  pages={100-114},
  doi={10.1109/MSP.2019.2913173}}

@ARTICLE{engels2021autoradar,
  author={Engels, Florian and Heidenreich, Philipp and Wintermantel, Markus and Stäcker, Lukas and Al Kadi, Muhammed and Zoubir, Abdelhak M.},
  journal={IEEE Journal of Selected Topics in Signal Processing}, 
  title={Automotive Radar Signal Processing: Research Directions and Practical Challenges}, 
  year={2021},
  volume={15},
  number={4},
  pages={865-878},
  doi={10.1109/JSTSP.2021.3063666}}

@ARTICLE{hakob2019ieeespsmagazine,
  author={Hakobyan, Gor and Yang, Bin},
  journal={IEEE Signal Processing Magazine}, 
  title={High-Performance Automotive Radar: A Review of Signal Processing Algorithms and Modulation Schemes}, 
  year={2019},
  volume={36},
  number={5},
  pages={32-44},
  doi={10.1109/MSP.2019.2911722}}

@ARTICLE{mansi2022spatialsensing,
  author={Gupta, M. and Sharma, S. and Joshi, H. and Darak, S. J.},
  journal={IEEE Transactions on Circuits and Systems II: Express Briefs}, 
  title={Reconfigurable Architecture for Spatial Sensing in Wideband Radio Front-End}, 
  year={2022},
  volume={69},
  number={3},
  pages={1054-1058},
  doi={10.1109/TCSII.2021.3124449}}

@ARTICLE{li2022musictcas1,
  author={Li, Zeying and Wang, Weijiang and Jiang, Rongkun and Ren, Shiwei and Wang, Xiaohua and Xue, Chengbo},
  journal={IEEE Transactions on Circuits and Systems I: Regular Papers}, 
  title={Hardware Acceleration of MUSIC Algorithm for Sparse Arrays and Uniform Linear Arrays}, 
  year={2022},
  volume={69},
  number={7},
  pages={2941-2954},
  doi={10.1109/TCSI.2022.3162303}}

@ARTICLE{sharma2024dlforce_tcas1,
  author={Sharma, Animesh and Haq, Syed Asrar Ul and Darak, Sumit J.},
  journal={IEEE Transactions on Circuits and Systems I: Regular Papers}, 
  title={Low Complexity Deep Learning Augmented Wireless Channel Estimation for Pilot-Based OFDM on Zynq System on Chip}, 
  year={2024},
  volume={71},
  number={5},
  pages={2334-2347},
  doi={10.1109/TCSI.2024.3371780}}

@ARTICLE{Drozdenko2018hwswzynq,
  author={Drozdenko, Benjamin and Zimmermann, Matthew and Dao, Tuan and Chowdhury, Kaushik and Leeser, Miriam},
  journal={IEEE Transactions on Emerging Topics in Computing}, 
  title={Hardware-Software Codesign of Wireless Transceivers on Zynq Heterogeneous Systems}, 
  year={2018},
  volume={6},
  number={4},
  pages={566-578},
  doi={10.1109/TETC.2017.2651054}}

@article{schweizer2021fairy,
  title={The fairy tale of simple all-digital radars: How to deal with 100 Gbit/s of a digital millimeter-wave MIMO radar on an FPGA [application notes]},
  author={Schweizer, Benedikt and Grathwohl, Alexander and Rossi, Gilberto and Hinz, Philipp and Knill, Christina and Stephany, Simon and Ng, Herman Jalli and Waldschmidt, Christian},
  journal={IEEE Microwave Magazine},
  volume={22},
  number={7},
  pages={66--76},
  year={2021},
  publisher={IEEE}
}

@INPROCEEDINGS{tewari2022reconfigofdm,
  author={Tewari, Aakanksha and Singh, Neelam and Darak, Sumit J. and Kizheppatt, Vipin and Jafri, Mohammed Sajjad},
  booktitle={2022 IEEE 65th International Midwest Symposium on Circuits and Systems (MWSCAS)}, 
  title={Reconfigurable Wireless PHY with Dynamically Controlled Out-of-Band Emission on Zynq SoC}, 
  year={2022},
  volume={},
  number={},
  pages={1-4},
  doi={10.1109/MWSCAS54063.2022.9859446}}

@ARTICLE{Lin2024HWspaceborneSAR,
  author={Lin, Jia-Zhao and Chen, Po-Ta and Chin, Hung-Yuan and Tsai, Pei-Yun and Lee, Sz-Yuan},
  journal={IEEE Transactions on Very Large Scale Integration (VLSI) Systems}, 
  title={Design and Implementation of a Real-Time Imaging Processor for Spaceborne Synthetic Aperture Radar With Configurability}, 
  year={2024},
  volume={32},
  number={4},
  pages={669-681},
  doi={10.1109/TVLSI.2023.3338476}}

@ARTICLE{zhang2023HW_EVD_tvt,
  author={Zhang, Xiao-Wei and Yan, Di and Zuo, Lei and Li, Ming and Guo, Jian-Xin},
  journal={IEEE Transactions on Vehicular Technology}, 
  title={High-Performance of Eigenvalue Decomposition on FPGA for the DOA Estimation}, 
  year={2023},
  volume={72},
  number={5},
  pages={5782-5797},
  doi={10.1109/TVT.2022.3221915}}

@ARTICLE{CAS1_beamtrain,
  author={Boljanovic, Veljko and Yan, Han and Lin, Chung-Ching and Mohapatra, Soumen and Heo, Deukhyoun and Gupta, Subhanshu and Cabric, Danijela},
  journal={IEEE Transactions on Circuits and Systems I: Regular Papers}, 
  title={Fast Beam Training With True-Time-Delay Arrays in Wideband Millimeter-Wave Systems}, 
  year={2021},
  volume={68},
  number={4},
  pages={1727-1739},
  doi={10.1109/TCSI.2021.3054428}}

@ARTICLE{CAS1_beamtrain2,
  author={Mirfarshbafan, Seyed Hadi and Gallyas-Sanhueza, Alexandra and Ghods, Ramina and Studer, Christoph},
  journal={IEEE Transactions on Circuits and Systems I: Regular Papers}, 
  title={Beamspace Channel Estimation for Massive MIMO mmWave Systems: Algorithm and VLSI Design}, 
  year={2020},
  volume={67},
  number={12},
  pages={5482-5495},
  doi={10.1109/TCSI.2020.3023023}}

@ARTICLE{CAS1_80211ad,
  author={Liu, Wei-Chang and Wei, Ting-Chen and Huang, Ya-Shiue and Chan, Ching-Da and Jou, Shyh-Jye},
  journal={IEEE Transactions on Circuits and Systems I: Regular Papers}, 
  title={All-Digital Synchronization for SC/OFDM Mode of IEEE 802.15.3c and IEEE 802.11ad}, 
  year={2015},
  volume={62},
  number={2},
  pages={545-553},
  doi={10.1109/TCSI.2014.2361035}}

@ARTICLE{RSP_hw1,
  author={Yeh, Chun-Yu and Chu, Ting-Chung and Chen, Chiao-En and Yang, Chia-Hsiang},
  journal={IEEE Transactions on Circuits and Systems I: Regular Papers}, 
  title={A Hardware-Scalable DSP Architecture for Beam Selection in mm-Wave MU-MIMO Systems}, 
  year={2018},
  volume={65},
  number={11},
  pages={3918-3928},
  doi={10.1109/TCSI.2018.2856124}}

@INPROCEEDINGS{RSP_hw2,
  author={Yuan, Quan and Zhuge, Shun and Lin, Zhiping and Ma, Yugang and Zeng, Yonghong},
  booktitle={2025 IEEE International Symposium on Circuits and Systems (ISCAS)}, 
  title={Kalman Filtering based Target Tracking for Multistatic Sensing in ISAC Systems}, 
  year={2025},
  volume={},
  number={},
  pages={1-5},
  doi={10.1109/ISCAS56072.2025.11043835}}

@ARTICLE{RSP_hw3,
  author={Xu, Xuanzhe and Liu, Xianjun and Wei, Siyuan and Yu, Shuangming and Dou, Runjiang and Yang, Xu and Wang, Zhe and Liu, Jian and Wu, Nanjian and Liu, Liyuan},
  journal={IEEE Transactions on Circuits and Systems I: Regular Papers}, 
  title={A 75.6 Gb/s 22-bit Floating-Point Coarse-Grained Versatile DSP Embedding 26K-Point Baseband Signal Processing and 2048 × 256-Point Complex FFT}, 
  year={2025},
  volume={},
  number={},
  pages={1-13},
  doi={10.1109/TCSI.2025.3578436}}

@INPROCEEDINGS{RSP_hw0,
  author={Yao, Long and Li, Yan and Li, Jianxin and Yu, Xinyuan and Wei, Ning},
  booktitle={2025 IEEE International Symposium on Circuits and Systems (ISCAS)}, 
  title={Efficient Radar Signal Processing in ISAC: Optimizing FFT Operations with Circulant Matrix}, 
  year={2025},
  volume={},
  number={},
  pages={1-5},
  doi={10.1109/ISCAS56072.2025.11043701}}

@ARTICLE{MUSIC_HW1,
  author={Butt, Uzma M. and Khan, Shoab A. and Ullah, Anees and Khaliq, Abdul and Reviriego, Pedro and Zahir, Ali},
  journal={IEEE Transactions on Circuits and Systems I: Regular Papers}, 
  title={Towards Low Latency and Resource-Efficient FPGA Implementations of the MUSIC Algorithm for Direction of Arrival Estimation}, 
  year={2021},
  volume={68},
  number={8},
  pages={3351-3362},
  doi={10.1109/TCSI.2021.3083280}}

@ARTICLE{MUSIC_HW2,
  author={Sangbone Assoa, Adou and Bhat, Ashwin and Ryu, Sigang and Raychowdhury, Arijit},
  journal={IEEE Transactions on Circuits and Systems I: Regular Papers}, 
  title={MDS-DOA: Fusing Model-Based and Data-Driven Approaches for Modular, Distributed, and Scalable Direction-of-Arrival Estimation}, 
  year={2025},
  volume={72},
  number={2},
  pages={941-952},
  doi={10.1109/TCSI.2024.3469928}}

@ARTICLE{RSP_hw4,
  author={Qian, Junhui and Tian, Fengchun and Zhang, Yisha and Jiang, Anyan},
  journal={IEEE Transactions on Circuits and Systems II: Express Briefs}, 
  title={Joint Design for Cooperative Radar and Communication Systems in Multi-Target Optimization}, 
  year={2022},
  volume={69},
  number={2},
  pages={614-618},
  doi={10.1109/TCSII.2021.3100350}}

@ARTICLE{ISAC_Ieee_proc,
  author={González-Prelcic, Nuria and Furkan Keskin, Musa and Kaltiokallio, Ossi and Valkama, Mikko and Dardari, Davide and Shen, Xiao and Shen, Yuan and Bayraktar, Murat and Wymeersch, Henk},
  journal={Proceedings of the IEEE}, 
  title={The Integrated Sensing and Communication Revolution for 6G: Vision, Techniques, and Applications}, 
  year={2024},
  volume={112},
  number={7},
  pages={676-723},
  doi={10.1109/JPROC.2024.3397609}}

@misc{TI_mmWaveRadarSensors,
  author = {Texas Instruments},
  title        = {Millimeter Wave Radar Sensors},
  howpublished = {\url{https://www.ti.com/product-category/sensors/mmwave-radar/products.html}},
  note         = {Accessed: Nov. 24, 2025}
}

@misc{Infineon_24GHzRadarIoT,
  title        = {24 GHz Radar Sensors for IoT},
  howpublished = {\url{https://www.infineon.com/products/sensor/radar-sensors/radar-sensors-for-iot/24ghz-radar}},
  note         = {Accessed: Nov. 24, 2025},
  author = {Infineon Technologies}
}

@misc{NXP_SAF85XXL,
  title        = {77 GHz RFCMOS Automotive Radar SoC},
  howpublished = {\url{https://www.nxp.com/products/radio-frequency-rf/radar-transceivers-and-socs/one-chip-lip-77ghz-rfcmos-automotive-radar-soc:SAF85XXL}},
  note         = {Accessed: Nov. 24, 2025},
  author = {NXP Semiconductors}
}

\end{document}